\pdfoutput=1
\documentclass[a4paper]{article}
\usepackage[margin=2.5cm]{geometry}
\usepackage[numbers]{natbib}
\usepackage[bookmarks, pagebackref=false]{hyperref}
\usepackage{doi}
\usepackage[usenames,dvipsnames]{xcolor}
\usepackage{authblk}
\usepackage{appendix}
\usepackage{indentfirst}

\usepackage{amsmath}
\usepackage{paralist}
\usepackage{bbold}

\usepackage{tikz}
\usetikzlibrary{intersections}
\usetikzlibrary{arrows,decorations.pathmorphing,backgrounds,positioning,fit,petri}
\usetikzlibrary{decorations.markings}
\usetikzlibrary{calc}
\usetikzlibrary{intersections,through,hobby}

\usepackage{cleveref}

\definecolor{bla}{HTML}{03396C}
\definecolor{blaa}{HTML}{005B96}
\definecolor{blaaa}{HTML}{6497B1}
\hypersetup{
  colorlinks, 
  bookmarksopen, 
  bookmarksnumbered,
  citecolor=blaa, 		
  linkcolor=blaa,    	
  urlcolor=blaa,			
}

\newcommand{\ep}{\ensuremath{\varepsilon}}
\newcommand{\MS}{\ensuremath{\overline{\rm MS}}}

\newcommand{\vphi}{\ensuremath{\varphi}}

\newcommand{\Zspp}{\ensuremath{Z_{\sigma\varphi\varphi}}}
\newcommand{\Zpp}{\ensuremath{Z_{\varphi\varphi}}}
\newcommand{\Zss}{\ensuremath{Z_{\sigma\sigma}}}
\newcommand{\Zsss}{\ensuremath{Z_{\sigma\sigma\sigma}}}
\newcommand{\Zp}{\ensuremath{Z_{\vphi}}}
\newcommand{\Zs}{\ensuremath{Z_{\sigma}}}
\newcommand{\ZgA}{\ensuremath{Z_{g_1}}}
\newcommand{\ZgB}{\ensuremath{Z_{g_2}}}

\newcommand{\K}{\ensuremath{\mathcal{K}}}                       

\newcommand{\Rp}{\ensuremath{\mathcal{R}^{\prime}}}
\newcommand{\Rs}{\ensuremath{\mathcal{R}^{*}}}
\newcommand{\KRp}{\ensuremath{\mathcal{K}\mathcal{R}^{\prime}}}
\newcommand{\Op}[1]{\ensuremath{\mathcal{O}_{#1}}}

\title{\color{bla} Critical exponents from five-loop scalar theory\\ renormalization near six-dimensions}
\author[1]{Mikhail Kompaniets}
\author[2]{Andrey Pikelner}
\affil[1]{ St. Petersburg State University,\protect\\
  7/9 Universitetskaya nab., St. Petersburg, 199034 Russia.}
\affil[2]{Bogoliubov Laboratory of Theoretical Physics,
  Joint Institute for Nuclear Research,\protect\\
  Joliot-Curie 6, Dubna 141980, Russia}
  
\date{}
\begin{document}

\maketitle
\begin{abstract}
  We present five-loop results for the renormalization of various models with a
  cubic interaction (in $d=6-2\ep$ dimensions). For the scalar model and its
  $O(n)$-symmetric extension we provide renormalization constants, anomalous
  dimensions and critical exponents. We discuss in detail the method of
  calculation, and provide all counterterms up to five loops. This allows one to
  consider generalizations of the $\vphi^3$ theory to other symmetries.
\end{abstract}

\allowdisplaybreaks

\section{Introduction}
\label{sec:intro}
In the quantum field theory there are two widely known models used as a
playground for new techniques of multi-loop calculations, these are models with
quadratic interaction ($\lambda\vphi^4$) and cubic interaction
($\lambda\vphi^3$). While these models are most simple among the non-trivial
theories, they admit a number of generalizations which are interesting from the
physical point of view. Most famous generalization is $O(n)$-symmetric models:
$O(n)$-symmetric $\lambda\vphi^4$ theory describes critical behavior of
different systems (among them are liquid-vapor transitions, superfluid
transition of helium, uniaxial magnets, isotropic ferromagnets etc.)
\cite{Pelissetto:2000ek,Vasilev,Kompaniets:2017yct}, while $O(n)$ $\vphi^3$
theory generalisation is widely known from \cite{Fei:2014yja,Fei:2014xta} as
candidate for model dual to the Vasiliev higher-spin theory in
AdS$_6$~\cite{Fradkin:1987ks,Vasiliev:2003ev}. Also many other generalizations
are available for these theories, among them we can mention
cubic~\cite{Adzhemyan:2019gvv} and chiral~\cite{Kompaniets:2019xez} $\vphi^4$
models which represent physicaly interesting universality classes laying outside
of $O(n)$ model. For $\vphi^3$ theory most famous generalization is the
$q$-state Potts model~\cite{Potts:1951rk} and its special case ($q=1$)
percolation
theory~\cite{deAlcantaraBonfim:1980pe,deAlcantaraBonfim:1981sy,Gracey:2015tta}.
Recently significant progress was made in $\lambda\vphi^4$ model:
six-~\cite{Batkovich:2016jus,Kompaniets:2016hct,Kompaniets:2017yct} and even
seven-loop~\cite{Schnetz:2016fhy} anomalous dimensions and critical exponents
were calculated. For the second model and its generalizations only four-loop
results are available~\cite{Gracey:2015tta} in analytical form and just recently
five-loop calculations were presented~\cite{BGS5L}. In this paper we present a
simple method to calculate renormalization group functions and critical
exponents in any generalization of the $\lambda \vphi^3$ theory in $d=6-2\ep$
dimensions up to five loops.
The paper is organized as follows: in section~\ref{sec:phi3RG} we describe the
technique used for five-loop calculations of scalar $\lambda \vphi^3$ model. In
section~\ref{sec:phi3On} using $O(n)$-symmetric model as an example we show how
to perform calculations in generalizations of the scalar $\lambda \vphi^3$ model
using results for counterterms for the individual diagrams available as a
byproduct of calculations in the model~\ref{sec:phi3RG}. For this model we
calculate critical dimensions of different operators allowing comparison with
$1/n$ expansion results available in the literature.
\section{Simple scalar model}
\label{sec:phi3RG}
We start with the 1-component scalar model in $d=6-2\ep$ dimensions, defined by the
bare Lagrangian
\begin{equation}
  \label{eq:lagPhi3}
  \mathcal{L}=  \frac{1}{2}(\partial_{\mu} \varphi_B)^2 + \frac{1}{2} m_B^2 \varphi_B^2 + g_B\frac{(4\pi)^3}{6} \varphi_B^3.
\end{equation}
Bare and renormalized parameters in the $\MS$ scheme are related by
\begin{equation}
  \label{eq:bare2ren}
  \vphi_B = \Zp \vphi, \quad g_B = \mu^{\ep} Z_g g, \quad m_B = Z_m m.
\end{equation}
For our first step to the five-loop renormalization-group functions,
we need to calculate the renormalization constants $Z_i$ defined in
eq.~\eqref{eq:bare2ren}. Unfortunately, it is very difficult to extend the
techniques used in Ref.~\cite{Gracey:2015tta} to the five-loop level.
Four-loop calculations~\cite{Gracey:2015tta} were performed with a relatively
straightforward approach, by transforming all diagrams to 4-loop massless
propagator type integrals. The such obtained integrals can then be reduced to a small set of
master integrals using IBP(integration by parts) relations, as four-loop IBP reduction is well
developed. Another important property of the four-loop propagator type integrals
reduction is that it is possible to derive dimensional recurrence relations
for master integrals in $d=6-2\ep$ from known $d=4-2\ep$
results\cite{Baikov:2010hf,Lee:2011jt}. Unfortunately, this approach cannot
directly be applied to
five-loop calculations, as there is no efficient way to reduce
five-loop massless propagator integrals to the master integrals. It is
especially difficult to derive dimensional recurrence relations to use the
known $d=4-2\ep$ five-loop master integrals~\cite{Georgoudis:2018olj} in
the $d=6-2\ep$ problem we are dealing with.
More sophisticated methods were applied in the breakthrough calculations in
$\lambda\vphi^4$ theory in
$d=4-2\ep$~\cite{Batkovich:2016jus,Kompaniets:2016hct,Kompaniets:2017yct,Schnetz:2016fhy}.
In Ref.~\cite{Batkovich:2016jus} anomalous dimension of the field was calculated
by using the infrared
$\Rs$-operation~\cite{ChetyrkinGorishnyLarinTkachov:Analytical5loop,Chetyrkin:1982nn,Chetyrkin:1984xa,Chetyrkin:1991mw,Chetyrkin:2017ppe}
and four-loop IBP reduction. Due to the specific structure of diagrams with only
quartic vertices, the $\Rs$-operation is extremely effective, and permits to
reduce each of the six-loop two-point diagrams to four-loop massless
propagators, which can then be further reduced to known master
integrals\cite{Baikov:2010hf,Lee:2011jt}. However, 12 non-primitive diagrams
(diagrams containing subdivergences) of the four-point function cannot be
calculated with the $\Rs$-operation and four-loop IPB reduction.
Therefore, a new method~\cite{Kompaniets:2016hct} using a distinct $\Rp$
operation in combination with special one-scale subtractions and parametric
integration of hyperlogarithms (\texttt{HyperInt} package)
\cite{Panzer:2013cha,Panzer:2014caa} was
developed~\cite{Kompaniets:2016hct,Kompaniets:2017yct}.
Seven-loop terms were calculated with an even more advanced \textit{graphical
functions} technique~\cite{Schnetz:2016fhy}. This most significant recent
progress in $\lambda \vphi^4$ theory was possible due to the development of new
powerful methods for multi-loop calculations. It should be noted, however, that
it is posible due to the relative simplicity of diagrams constructed from
quartic vertices only.
The methods mentioned above can be applied to $\lambda\vphi^3$ theory as well:
graphical functions were applied recently~\cite{BGS5L} to the problem considered
in the present paper. Parametric integration using hyperlogarithms also works,
but the straightforward approach used
in~\cite{Kompaniets:2016hct,Kompaniets:2017yct} meets performance problems due
to the large number of integrations in parametric space, and the reach
combinatorial structure of diagrams in the $\lambda\vphi^3$ theory.
Recently, five-loop QCD renormalization was
performed~\cite{Baikov:2016tgj,Herzog:2017ohr,Luthe:2017ttc,Chetyrkin:2017bjc}.
Methods utilized there are much closer to the problem we are interested in here.
One idea is to avoid calculation of five-loop integrals, by reducing the problem
to the calculation of four-loop massless propagator integrals through
application of the $\Rs$-operation to individual diagrams~\cite{Herzog:2017ohr}
or applying the $\Rs$-operation in a global
form~\cite{Baikov:2016tgj,Chetyrkin:2017bjc}. The rest utilizes fully massive
five-loop tadpoles calculated numerically with high
precision~\cite{Luthe:2017ttc} at intermediate steps, and reconstructs the
analytical answer in the end.
Our starting point was to use the $\Rs$ approach, as it guaranties reduction of the
$L$-loop problem to a $(L-1)$-loop problem, in our case the reduction from five
to four loops, where the latter can be solved using available IBP reduction
tools. After a detailed analysis we realized that it is even possible to perform
calculations without the
$\Rs$-operation. Our strategy lies somewhere in between the
two methods used in the QCD calculations mentioned above. Similar to~\cite{Luthe:2017ttc}
we reduce the problem to the calculation of the divergent parts of specially
constructed fully massive five-loop tadpole integrals. But instead of
calculating integrals directly as in~\cite{Luthe:2017ttc}, we calculate their
divergences with the help of the \emph{infra-red rearrangement} (IRR)
trick, which requires knowledge of four-loop integrals only. This IRR technique is
valid due to the independence of the renormalization constants $Z_i$ of masses and
external momenta~\cite{Vladimirov:1979zm}.
\subsection{Details of calculation}
\label{sec:calcPhi3}

Due to the moderate number of diagrams in model
\eqref{eq:lagPhi3}, we apply the $\Rp$-operation separately to each diagram.
This approach allows us to study generalizations of
$\lambda\vphi^3$ theory to different symmetries or field contents, by simply
changing the symmetry factor of each diagram.
The $\Rp$-operation, also known as the incomplete $\mathcal{R}$-operation, subtracts
all subdivergences from a diagram, except the superficial divergence of the
diagram itself. Working in the $\MS$ scheme, it is useful to introduce the
$\mathcal{K}$ operation, which extracts the divergent part of the Laurent series
expansion in $\ep$. For each diagram $G$, the result $\KRp G$
gives the contribution to the corresponding renormalization constant in the $\MS$
scheme. Its recursive definition has the form
\begin{equation}
  \label{eq:RopDef}
  \KRp G = \mathcal{K} G - \mathcal{K}(1-\mathcal{R}^\prime) G
  = \mathcal{K} G + \sum\limits_{\{\gamma\}} \mathcal{K}\big[\prod\limits_{\gamma_i \in \{\gamma\}}(- \KRp \gamma_i) \ast G/\{\gamma\} \big].
\end{equation}
The sum is taken over all non-empty subsets $\{\gamma\}$ of non-intersecting
divergent subgraphs $\gamma_i$. The product is taken over all subgraphs in each
subset. The recursion stops when no $\{\gamma\}$ contains a diagram with
subdivergences. In this case, $\KRp \gamma = \mathcal{K}\gamma$ (pole part
of the diagram $\gamma$). The operation $\ast$ inserts the appropriate counterterm into
the co-graph $G/\{\gamma\}$. For a logarithmically divergent subgraph, it
reduces to a scalar multiplication, while for a quadratically divergent
subgraph $\gamma_i$ it inserts the full(momentum) structure $A p^2 + B m^2$.
In equation~\eqref{eq:RopDef}, we introduce the formal operation $(1-\mathcal{R}^\prime) G$,
which for an $L$-loop diagram $G$ contains only terms with less than
$L$ loops. Assuming our ability to calculate arbitrary loop integrals at level
$(L-1)$, such a splitting allows us to separate the complicated part containing $L$-loop
integrals ($\mathcal{K}G$ in ~\eqref{eq:RopDef}) from the rest.
The renormalization constants $Z_\vphi$, $Z_g$ and $Z_m$ are finally calculated from the following Green functions with all subdivergences
subtracted,
\begin{equation}
  \label{eq:phi3ZKRp1}
  Z_{\vphi}^2 = 1 + \KRp \partial_{q^2}\Gamma_{\vphi\vphi},
  \quad
  Z_{m}^2 Z_{\vphi}^2= 1 +  \KRp \partial_{m^2}\Gamma_{\vphi\vphi},
  \quad
  Z_{g} Z_{\vphi}^3 = 1 - \KRp \overline\Gamma_{\vphi\vphi\vphi},
  \quad
  \overline{\Gamma}_{\vphi\vphi\vphi} = \Gamma_{\vphi\vphi\vphi}/(-g\mu^{\varepsilon}).
\end{equation}
In order to realize this program, we need to compute the divergent parts of
diagrams contributing to the three-point function,  and derivatives (with respect to external momenta and mass) of
the two-point function. Since we are working in a scalar theory, integrals
with numerators(all propagators give denominators) only appear in the calculation of $Z_{\vphi}$ after
differentiation with respect to $q^2$. Other renormalization constants are calculated
using scalar integrals only. The most complicated part in the calculations of the
renormalization constants~\eqref{eq:phi3ZKRp1} is to compute the divergent part of
the five-loop diagrams. The independence of the $\MS$ renormalization constants of masses
and external momenta allows us to set to zero all masses and external momenta of
the original diagram, and to introduce artificial external momenta flowing through
a single edge,
\begin{equation}
  \label{eq:IRR5l}
  \KRp \left[\vcenter{\hbox{
        \begin{tikzpicture}[use Hobby shortcut, scale=0.7]
          \filldraw[color=gray!60, fill=gray!5, very thick] (0,0) circle (1);
          \draw[very thick] (90:1) -- (90:1.5);
          \draw[very thick] (210:1) -- (210:1.5);
          \draw[very thick] (330:1) -- (330:1.5);
          \fill (90:1) circle (2pt);
          \fill (210:1) circle (2pt);
          \fill (330:1) circle (2pt);
          \node[anchor=center] at (0,0) {\scriptsize $L=5$};
        \end{tikzpicture}
      }}\right]
  =
  \KRp \left[\vcenter{\hbox{
        \begin{tikzpicture}[use Hobby shortcut, scale=0.7]
          \draw[very thick] (180:1) arc (180:360:1);
          \draw[very thick] (180:1) -- (180:1.5);
          \draw[very thick] (0:1) -- (0:1.5);
          \filldraw[color=gray!60, fill=gray!5, very thick] (0:1) arc (0:180:1);
          \filldraw[color=gray!60, fill=gray!5, very thick] (180:1) .. (0,-0.4) .. (0:1);
          \fill (0:1) circle (2pt);
          \fill (90:1) circle (2pt);
          \fill (180:1) circle (2pt);
          \node[anchor=center] at (0,0.2) {\scriptsize $L=4$};
        \end{tikzpicture}
      }}\right].
\end{equation}
The result of the $\KRp$ operation does not change, if such a transformation is
applied to logarithmically divergent diagrams, and no IR divergences appear
after rearrangement. We have checked explicitly that for all three-point
integrals at least one such edge can be identified. This edge assigns external
momenta routing. As we can see in equation~\eqref{eq:IRR5l}, the rearranged
diagram has the form of a four-loop propagator inserted into a one-loop
propagator, and can easily be calculated since four-loop propagators are known.
The same technique is appropriate for the calculation of $Z_m$, since
differentiation with respect to $m^2$ is equal to the attachment of an
additional leg to one of the edges of the original two-point diagram.
The calculation of the field renormalization constant $Z_{\vphi}$ is more
complicated than that of $Z_m$, since differentiation with respect to the  external momenta produces
a large number of integrals with numerators, and it becomes difficult to apply
the IRR
trick in the form given in equation~\eqref{eq:IRR5l}. To overcome this difficulty, we developed
another approach, reducing the problem to the calculation of fully massive
tadpoles instead of massless propagators. This approach is applicable to the calculation of
all renormalization constants of interest. Our starting point is to split
the $\KRp G$ operation acting on a $L$-loop diagram $G$ into the two
parts given in equation~\eqref{eq:RopDef}. The first term $\mathcal{K} G$
contains divergences of $L$-loop diagrams, while the second term
$\mathcal{K}(1-\mathcal{R}^\prime) G$ contains subtraction terms with maximal
loop order $(L-1)$. Assuming that we know how to calculate an arbitrary integral
at loop order $(L-1)$ present in $\mathcal{K}(1-\mathcal{R}^\prime) G$, the main difficulty is to
calculate the $L$-loop part $\mathcal{K} G$.
To overcome the most complicated part, namely evaluating the poles of the
$L$-loop diagrams $\mathcal{K}G$, we consider divergences of the related
integrals instead. In comparison to diagrams, integrals satisfy more relations
allowing to reduce their complexity. One class of relations is \emph{Integration
By Parts}(IBP)~\cite{Tkachov:1981wb,Chetyrkin:1981qh}, widely used for the
reduction of large numbers of integrals to a smaller set of \emph{master
integrals}(MI).
IBP relations connect $m$ original integrals $J_a (a=1,\dots,m)$ with another
set of $n$ ($n \le m$) integrals $I_b (b=1,\dots, n)$ through a matrix $M_{ab}$
with rational polynomial coefficients in masses, scalar products of external
momenta and space-time dimension $d=6-2\ep$. For single-scale integrals it
simplifies to
\begin{equation}
  \label{eq:IBPrel} J_a(\ep) = M_{ab}(\ep) I_b(\ep)
\end{equation}
For arbitrary choice of integrals $I_b$, the matrix $M_{ab}(\ep)$ may contain
poles in $\ep$, and the calculation of $\mathcal{K} J_a$ require not only pole
parts for some integrals from $I_b$, but also their finite parts, and even
higher-order terms in the $\ep$-expansion. To avoid such complications, we focus
on candidate integrals $I_b$, for which the matrix $M_{ab}$ is regular in the
limit of $\ep \to 0$. A basis of master integrals with such properties was
considered in Ref.~\cite{Chetyrkin:2006dh} and called \textit{$\ep$-finite}.
Since we decided to reduce the problem to the calculation of $\KRp$ for fully
massive tadpole diagrams, we collect into the set $J_a$ all integrals
originated from vertex and propagator diagrams. In three-point integrals, we set to zero all
external momenta while treating all internal lines as massive. The same
procedure is used for two-point functions, after
differentiation in squared external momenta. Such an IR rearrangement is allowed,
since all integrals are logarithmically divergent, IR
divergences are regularized by masses on all internal lines, and the action of the $\KRp$
operation is independent of external momenta. As a result, all integrals $J_a$ are fully
massive tadpoles, possibly with numerators, and including all 5-loop integrals needed for
the $\mathcal{K}G$ operation in equation~\eqref{eq:phi3ZKRp1}. Integrals in the set
$I_b$ satisfy the conditions:
\begin{enumerate}
\item \label{itm:epMab} the matrix $M_{ab}$ does not have poles in $\ep$,
\item \label{itm:irr5l} $\mathcal{K} I_b$ can be calculated
  using four-loop integrals only using IRR trick,
\item \label{itm:intdia} the integrals $I_b$ do not have numerators and allow an
interpretation as diagrams in scalar field theory.
\end{enumerate}
Since we only need the divergent parts of the integrals $I_b$ (due to condition
\ref{itm:epMab}), we choose candidates for $I_b$ computable from four-loop
integrals(condition \ref{itm:irr5l}). Calculation of the divergent part using
IRR is based on the independence of counterterms of the logarithmic diagrams on
the infrared regularization \eqref{eq:IRR5l}. It can be rewritten as
\begin{equation}
  \label{eq:IRR5l2}
  \K \left[\vcenter{\hbox{
        \begin{tikzpicture}[use Hobby shortcut, scale=0.7]
          \filldraw[color=gray!60, fill=gray!5, very thick] (0,0) circle (1);
          \node[anchor=center] at (0,0) {\scriptsize $L=5$};
        \end{tikzpicture}
      }}\right]_{m\neq 0}
  =
  \KRp \left[\vcenter{\hbox{
        \begin{tikzpicture}[use Hobby shortcut, scale=0.7]
          \draw[very thick] (180:1) arc (180:360:1);
          \draw[very thick] (180:1) -- (180:1.5);
          \draw[very thick] (0:1) -- (0:1.5);
          \filldraw[color=gray!60, fill=gray!5, very thick] (0:1) arc (0:180:1);
          \filldraw[color=gray!60, fill=gray!5, very thick] (180:1) .. (0,-0.4) .. (0:1);
          \fill (0:1) circle (2pt);
          \fill (180:1) circle (2pt);
          \node[anchor=center] at (0,0.2) {\scriptsize $L=4$};
        \end{tikzpicture}
      }}\right]_{m = 0}- \K(1-\Rp) \left[\vcenter{\hbox{
        \begin{tikzpicture}[use Hobby shortcut, scale=0.7]
          \filldraw[color=gray!60, fill=gray!5, very thick] (0,0) circle (1);
          \node[anchor=center] at (0,0) {\scriptsize $L=5$};
        \end{tikzpicture}
      }}\right]_{m\neq 0},
\end{equation}
where both terms on the right-hand-side can be calculated using integrals with
four loops or less only.
The reduction of five-loop tadpole integrals is the most time-consuming part of
the calculation, and our choice of candidate integrals for the set $I_b$ is
limited. Therefore we try to include in the candidate list as few integrals as
possible, and mostly reuse results of IBP reduction for integrals
$J_a$(notations in equation~\eqref{eq:IBPrel}). A natural choice fulfilling all
three conditions is the set $J^{(3)}$ of the integrals needed for the
three-point function. As discussed before (condition~\ref{itm:irr5l}) their
divergences can be calculated using equation~\eqref{eq:IRR5l}, they are free
from numerators, and are in one-to-one correspondence with scalar theory
diagrams(condition~\ref{itm:intdia}). Unfortunately, it is not enough to include
integrals from the set $J^{(3)}$ satisfying condition~\ref{itm:epMab} and we
need to extend the list of candidates. Our choice is to include integrals
$J^{(4)}$ resulting from four-point functions in
$\vphi^3$-theory~\eqref{eq:lagPhi3}. Formally, the IRR trick~\eqref{eq:IRR5l2}
cannot be applied to such diagrams (as they are not logarithmically divergent),
but due to the absence of superficial divergences ($\KRp$ of such diagrams is
always zero) it is still possible to calculate their divergent parts from
knowledge of lower-loop integrals only,
\begin{equation}
  \label{eq:finKfromKRp}
  \mathcal{K}J_k^{(4)} = \underbrace{\KRp J_k^{(4)}}_{=0} + \mathcal{K}(1-\mathcal{R}^\prime)J_k^{(4)}.
\end{equation}
Combining integrals from $J^{(3)}$ with integrals from $J^{(4)}$ we are able to
construct the set of integrals $I_b$ meeting all three conditions. Trying to
maximize the number of integrals from the set $J^{(4)}$, only 14 integrals from
the set $J^{(3)}$ need to be calculated. From these 14 integrals 8 are free from
subdivergences, thereby the $\Rp$ operation acts trivially on them, and their
divergences can be calculated from the massless diagrams via
rule~\eqref{eq:IRR5l}. For the remaining 6 diagrams, where knowledge of the
$\KRp$ operation is required, details of the $\KRp$ calculation using the IRR
trick are presented in Appendix~\ref{sec:manual-dias}.
As an illustration, let us apply the described procedure to the most complicated
class of diagrams,
\begin{equation}
  \label{eq:intProp2HHHH}
  \KRp \partial_{q^2} J_i
  = \mathcal{K} \partial_{q^2} J_i - \mathcal{K}(1-\mathcal{R}^\prime) \partial_{q^2} J_i
  = \mathcal{K}\left[A_{ik}(\ep) \mathcal{K}J_k^{(3)} + B_{ik}(\ep) \mathcal{K}J_k^{(4)}\right]
  - \partial_{q^2}\mathcal{K}(1-\mathcal{R}^\prime)J_i.
\end{equation}
By using $\KRp \partial_{q^2} G = \partial_{q^2} \KRp G $, we interchanged the
order of differentiation in external momenta, and divergences subtraction, in
the last term. This allows us to get rid of the $\KRp$ operation applied to
diagrams with numerators in all steps of the calculation. The matrices
$A_{ik}(\ep)$ and $B_{ik}(\ep)$ produced as a result of the IBP reduction in
equation~\eqref{eq:intProp2HHHH} are free from poles in $\ep$.
As a result, we have the following workflow to calculate all renormalization
constants~\eqref{eq:phi3ZKRp1} up to five-loop order. First we generate all
three-point and two-point diagrams in scalar theory using
\texttt{DIANA}~\cite{Tentyukov:1999is}. At this step we assign equal masses to
all internal lines. For logarithmically divergent three-point functions we set
all external momenta to zero, for two-point functions we keep the routing of
external momenta. To each scalar diagram we apply the $\KRp$ operation,
implemented as a private \texttt{C++} code generating the appropriate
\texttt{FORM}~\cite{Kuipers:2012rf} input for the calculation of divergent
subgraphs and co-subgraphs according to equation~\eqref{eq:RopDef}. At this
step, co-subgraphs needed for the calculation of $\partial_{q^2}\KRp$ are
differentiated in external momenta, and external momenta are set to zero,
reducing the problem to the calculation of fully massive tadpoles only. All
required integrals up to three-loop order are calculated with the help of the
\texttt{MATAD-ng}\footnote{Available from
\url{https://github.com/apik/matad-ng}} version of the \texttt{MATAD}
package\cite{Steinhauser:2000ry}, operating in arbitrary space-time dimension
$d$, and at four-loop order using the package
\texttt{FMFT}\cite{Pikelner:2017tgv}. With the same technique, using
relation~\eqref{eq:finKfromKRp}, we obtain the divergences of all four-point
integrals $\mathcal{K}J^{(4)}$ from equation~\eqref{eq:intProp2HHHH}. Another
ingredient included in equation~\eqref{eq:intProp2HHHH} are divergences of the
three-point functions $\mathcal{K}J^{(3)}$, which, with the help of IRR, are
reduced to the calculation of the $\KRp$ operation on properly constructed
massless five-loop propagator diagram, in accordance with
equation~\eqref{eq:IRR5l}. Details of the calculation for the six diagrams with
subdivergences from the set $\mathcal{K}J^{(3)}$ are presented in
Appendix~\ref{sec:manual-dias}. Four-loop massless propagator integrals are
calculated with the package \texttt{FORCER}~\cite{Ruijl:2017cxj}, and at lower
loop orders with the \texttt{MINCER}\footnote{``Mincer exact'' package available
from \url{https://www.nikhef.nl/~form/maindir/packages/mincer/mincerex.tgz}}
package~\cite{Gorishnii:1989gt,Larin:1991fz}.
As a last step to apply \eqref{eq:intProp2HHHH}, we need to reduce five-loop
tadpole integrals to the minimal set and construct matrices $A_{ik}(\ep)$ and
$B_{ik}(\ep)$. To perform this reduction, we use the Laporta
algorithm~\cite{Laporta:2001dd} implemented in the recent \texttt{C++} version
of the \texttt{FIRE6} package~\cite{Smirnov:2019qkx}. To overcome the appearance
of lengthy $d$-dependent coefficients in front of integrals during the reduction
process, we activated the \texttt{FIRE6} reduction functionality, using modular
arithmetic for the polynomial algebra at intermediate steps. Symmetry relations
among different integrals used in the IBP reduction process were provided by the
\texttt{LiteRed}~\cite{Lee:2012cn} package. To check divergent parts of
five-loop tadpoles numerically, we use the code
\texttt{FIESTA}~\cite{Smirnov:2015mct}.
All our results for the divergent parts of 3-point and 4-point integrals, of
$\KRp$ applied to each of the three-point diagrams, and of $\KRp \partial_{q^2}$
and $\KRp \partial_{m^2}$ applied to each of the two-point diagrams are
available as ancillary files with the arXiv version of the paper.
\subsection{Results for the scalar model}
\label{sec:phi3res}
From the renormalization constants in equation~\eqref{eq:bare2ren} we extract
the renormalization group functions
\begin{equation}
  \label{eq:betaPhi3}
  \beta_{g} = -\frac{\ep g}{1 + g \partial_g \log{Z_g}}, \quad
  \gamma_{\vphi} = \beta_g\;\partial_g \ln{Z_\vphi} , \quad
  \gamma_{m} = \beta_g\;\partial_g \ln{Z_{m}}.
\end{equation}
While in the above expressions the renormalization constants $Z_i$ have poles,
the final resulrs are free from poles, which is a strong check for the
correctness of our calculation. Expressions for anomalous dimensions
\eqref{eq:gamPhiPhi3} and \eqref{eq:gamMMPhi3} are provided in appendix
\ref{sec:AnDimPhi3}.  The $\beta$ function is related to the above results via
$\beta_g = -\ep g + g \left(\gamma_{\vphi} - 2\gamma_{m}\right)$. From
$\beta_g(g^*) = 0$ we can find the position of the fixed point $g^*$.
Substituting the value for $g^*$ into the anomalous dimension of the field, we obtain the
critical exponent for the dimension of the field,
\begin{equation}
  \label{eq:etaPhi3}
  \eta = 2 \gamma_{\vphi}^* = -0.222222 \,\ep - 0.23594 \,\ep^2 + 0.34952 \,\ep^3
  - 1.26327 \,\ep^4 + 6.66413 \,\ep^5 + O\left(\ep^6\right).
\end{equation}
Due to the relation $2\gamma_{m} = \gamma_{\vphi} - \left(\frac{\beta}{g} +
\ep\right)$ between anomalous dimensions and the $\beta$ function in the
model~\eqref{eq:lagPhi3}, another critical exponent, corresponding to the critical
dimension of the dimension-two operator, is trivially connected with
equation~\eqref{eq:etaPhi3} through the relation
\begin{equation}
  \label{eq:invNuPhi3}
  \frac{1}{\nu} = 2 + 2\gamma_{m}^* = 2 - \ep + \gamma_{\vphi}^* = 2 - \ep + \frac{\eta}{2}.
\end{equation}
For the correction-to-scaling exponent $\omega$ we have 
\begin{equation}
  \label{eq:OmegaPhi3}
  \omega = \left(\frac{\partial \beta}{\partial g}\right)_{g=g_*} = 2 \ep - 3.08642 \ep^2 + 12.7253 \ep^3 - 72.522 \ep^4 + 493.942 \ep^5 + O\left(\ep^6\right).
\end{equation}
The value found in equation~\eqref{eq:etaPhi3} is in agreement with the
four-loop results~\cite{2011TMP...169.1450A,Pismenskii:2015xxg,Gracey:2015tta}
and with the recent five-loop results obtained using the graphical-functions
technique~\cite{BGS5L}. With the first five terms of
expansion~\eqref{eq:etaPhi3} at hand, we can compare them to predictions for the
asymptotic behavior of higher-order terms derived in
Ref.~\cite{Kalagov:2014esa}:
\begin{equation}
  \label{eq:etaAsympPhi3}
  \eta_{\rm as} = \sum\limits_{n=1}^{\infty} f_n \ep^n, \quad f_n = -0.000586 \cdot  \sqrt{2\pi }\left(-\frac{5}{9}\cdot\frac{n}{e}\right)^n n^{5}\left(1 + O\left(\frac{1}{n}\right)\right).
\end{equation}
Comparing the asymptotic expansion~\eqref{eq:etaAsympPhi3} with
equation~\eqref{eq:etaPhi3}, we see that the low-order difference is huge($f_3 =
0.0822729$ and $f_4 = -0.67183$) but becomes smaller for the five-loop term($f_5
= 5.11508$). It will be interesting to compare with six-loop results once they
become available.
\section{Model with $O(n)$ symmetry and $1/n$ expansion}
\label{sec:phi3On}
As previously mentioned, one can use the results of section~\ref{sec:calcPhi3}
for calculations in more advanced theories. In this section we focus on the
$O(n)$-symmetric generalization of the model~\eqref{eq:lagPhi3}. Such a model
was considered in Ref.~\cite{Fei:2014yja}, and is interesting for the
investigation of the universality class with $O(n)$-symmetry above four
dimensions. Another important property of $O(n)$ symmetric models is that
critical exponents of the corresponding universality class can be calculated in
the framework of a $1/n$ expansion for arbitrary space-time dimension, thus
providing an additional check for the results obtained.
The bare Lagrangian of the model has the form
\begin{equation}
  \label{eq:lagPhi3On} \mathcal{L}= \frac{1}{2}(\partial_\mu \varphi_{i,B})^2 +
\frac{1}{2}(\partial_{\mu} \sigma_{B})^2 + g_{1,B}\frac{(4\pi)^3}{2}\sigma_{B}
(\varphi_{i,B})^2 + g_{2,B}\frac{(4\pi)^3}{6}\sigma_B^3,
\end{equation}
where field $\vphi_i$ is an $n$-component vector, and $\sigma$ is a scalar. We
start from the massless model and extract anomalous dimensions for dimension-two
operators. This is in contrast to the model~\eqref{eq:lagPhi3} where we
calculated the anomalous dimension of the mass in a theory where the mass term
is in the Lagrangian. Anomalous dimensions are now calculated from diagrams with
$\vphi^2$ and $\sigma^2$ insertions. Similar to equation~\eqref{eq:bare2ren} we
define renormalization constants connecting bare and renormalized parameters,
\begin{equation}
  \label{eq:ZdefOnRenBare}
  \vphi_{i,B} = \Zp \vphi_i, \quad \sigma_B = \Zs \sigma, \quad g_{1,B} = \mu^{\ep} Z_{g_1} g_1 , \quad g_{2,B} = \mu^{\ep} Z_{g_2} g_2.
\end{equation}
Taking into account symmetry coefficients and group-theory factors for the
model~\eqref{eq:lagPhi3On} and utilizing $\KRp$-operation results from the
section~\ref{sec:calcPhi3}, we obtain renormalization constants for the
following Green functions
\begin{equation}
  \label{eq:Zdias}
  \Zpp = 1 + \partial_{q^2} \KRp \Gamma_{\vphi\vphi}, \quad
  \Zss = 1 + \partial_{q^2} \KRp \Gamma_{\sigma\sigma}, \quad
  \Zspp = 1 - \KRp \overline\Gamma_{\sigma\vphi\vphi}, \quad
  \Zsss = 1 - \KRp \overline\Gamma_{\sigma\sigma\sigma}.
\end{equation}
Here $\overline\Gamma_{\sigma\vphi\vphi} =
\Gamma_{\sigma\vphi\vphi}/(-g_1\mu^\ep)$ and
$\overline\Gamma_{\sigma\sigma\sigma} =
\Gamma_{\sigma\sigma\sigma}/(-g_2\mu^\ep)$. From these functions, the
renormalization constants~\eqref{eq:ZdefOnRenBare} are extracted via the
relations
\begin{equation}
  \label{eq:ZdefOn}
  \Zp = \sqrt{\Zpp}, \quad \Zs = \sqrt{\Zss}, \quad \ZgA = \frac{\Zspp}{\Zs\Zp^2}, \quad   \ZgB = \frac{\Zsss}{\Zs^3}.
\end{equation}
Charge renormalization constants $\ZgA$ and $\ZgB$ provide us with
beta-functions for the two coupling constants of the theory,
\begin{align}
  \beta_{g_1} & = \frac{\partial g_1}{\partial \ln{\mu}} =  \frac{\ep g_1\left(1 + g_2 \partial_{g_2}\ln{\ZgB} - g_2 \partial_{g_2}\ln{\ZgA}  \right)}
                {g_1 g_2 \left(\partial_{g_2}\ln{\ZgA}\right) \left(\partial_{g_1}\ln{\ZgB}\right)
                -(1 + g_1 \partial_{g_1} \ln{\ZgA})(1 + g_2 \partial_{g_2} \ln{\ZgB})}
                \nonumber\\
  \beta_{g_2} & = \frac{\partial g_2}{\partial \ln{\mu}}  = \frac{\ep g_2\left( 1 + g_1  \partial_{g_1}\ln{\ZgA} - g_1 \partial_{g_1}\ln{\ZgB}  \right)}
                {g_1 g_2 \left(\partial_{g_2}\ln{\ZgA}\right) \left(\partial_{g_1}\ln{\ZgB}\right)
                - (1 + g_1 \partial_{g_1} \ln{\ZgA})(1 + g_2 \partial_{g_2} \ln{\ZgB})}  \label{eq:betaDefPoles}
\end{align}
Renormalization constants for the field yield the anomalous field dimensions
\begin{equation}
  \label{eq:gammaDefOn}
  \gamma_{\vphi} = \beta_{g_1}\;\partial_{g_1}\ln{Z_{\vphi}} + \beta_{g_2}\;\partial_{g_2} \ln{Z_{\vphi}},
  \quad
  \gamma_{\sigma} = \beta_{g_1}\;\partial_{g_1}\ln{Z_{\sigma}}  + \beta_{g_2}\;\partial_{g_2}\ln{Z_{\sigma}}.
\end{equation}
As in the previous section, cancellation of poles in
equations~\eqref{eq:betaDefPoles} and~\eqref{eq:gammaDefOn} is a strong test of
our calculations. Results for the beta-functions and anomalous dimensions are
quite lengthy to present here, but are available in the supplementary material
to the paper. They are in agreement with existing three-loop~\cite{Fei:2014xta}
and four-loop~\cite{Gracey:2015tta} calculations.
With results for the RG-functions at hand, we proceed with the analysis of the
critical behavior of the model~\eqref{eq:lagPhi3On}: Zeroes of the
beta-functions~\eqref{eq:betaDefPoles} $\beta_1 = \beta_2 = 0$ define multiple
fixed points $(g_1^*,g_2^*)$. We choose the unique IR-stable one corresponding
to the large-$n$ solution considered in Ref.~\cite{Fei:2014yja}.
Since our goal is to compare results of our calculation with available large-$n$
expansions for the critical exponents, we make an ansatz for the solution of the
fixed point $(g_1^*,g_2^*)$ as a double expansion in $\ep$ and $1/n$. Expansions
up to $O(1/n^4)$ and up to five loops, i.e. $\mathcal{O}(\ep^5)$, are presented
in equation~\eqref{eq:g1crit} and~\eqref{eq:g2crit}. Substituting the found
large-$n$ solution for the critical point as a $1/n$ expansion into the
anomalous field dimension~\eqref{eq:gammaDefOn}, we obtain the critical
dimensions of operators $\Delta_{\vphi}$ and $\Delta_\sigma$ as
\begin{equation}
  \label{eq:DeltaPhiSigma}
  \Delta_\vphi = \frac{d}{2}-1 + \gamma_\vphi^* = \frac{d}{2}-1 + \frac{\eta}{2},
  \quad
  \Delta_\sigma = \frac{d}{2}-1 + \gamma_\sigma^* =d-\frac{1}{\nu}.
\end{equation}
Results for $\eta$ up to $O(1/n^3)$ can be found in Ref.~\cite{Vasiliev:1982dc}
and for $1/\nu$ up to order $O(1/n^2)$ in~\cite{Vasiliev:1981dg}. Using the
expressions for $\gamma_{\vphi}^*$ and $\gamma_{\sigma}^*$, and the results for
the critical exponents, the critical dimensions as given in
equation~\eqref{eq:DeltaPhiSigma} are equal to all known orders in $1/n$
expansion.
In addition to the critical dimensions of fields~\eqref{eq:DeltaPhiSigma}, we
consider critical dimensions of operators with canonical dimension two. The main
complication in this case is operator mixing under the renormalization
procedure. We start with the definition of two operators with equal canonical
dimension,
\begin{equation}
  \label{eq:Opdef}
  \Op{1} = \vphi_i\vphi_i \quad \Op{2} = \sigma^2.
\end{equation}
The renormalization of the operators is defined as
\begin{equation}
    Z^{\Op{}}_{ij} \Op{j}^R = \Op{i}.
\end{equation}
The matrix of the renormalization constants $Z=(Z^{\Op{}})^{-1}$ can be
determined from the set of one-particle irreducible(1PI) Green functions with
operator insertions $\{\left<\mathcal{O}_1 \vphi_i \vphi_i \right>,
\left<\mathcal{O}_1 \sigma \sigma \right>, \left<\mathcal{O}_2 \vphi_i \vphi_i
\right>,\left<\mathcal{O}_2 \sigma \sigma \right>\}$. As in the previous
sections, we subtract all subdivergences with the help of the $\KRp$-operation,
\begin{equation}
  \label{eq:ZijDef}
  Z=\left(
    \begin{matrix}
      (1-\KRp \left<\mathcal{O}_1 \vphi_i \vphi_i \right>) Z_{\vphi}^{-2}\quad
      & -\KRp \left<\mathcal{O}_1 \sigma \sigma\right> Z_{\sigma}^{-2}\\
      -\KRp \left<\mathcal{O}_2 \vphi_i\vphi_i \right> Z_{\vphi}^{-2}\quad
      & (1-\KRp \left<\mathcal{O}_2 \sigma\sigma \right>) Z_{\sigma}^{-2}
    \end{matrix}\right).
\end{equation}
The anomalous dimensions matrix is then defined as
\begin{equation}
  \gamma^{\Op{}}_{ij} =(Z^{\Op{}})^{-1}_{ik} \left[\beta_1 \partial_{g_1}  + \beta_2 \partial_{g_2} \right] Z^{\Op{}}_{kj}
  = -\left( \left[\beta_1 \partial_{g_1}  + \beta_2 \partial_{g_2} \right] Z_{ik} \right) Z^{\Op{}}_{kj} .
  \label{eq:gammaO}
\end{equation}
Five-loop results for the matrix elements $\gamma_{ij}$ are attached to the
paper in a computer readable form, and the four-loop part is in agreement with
previous calculations~\cite{Gracey:2015tta}.
Critical dimensions of the mixing operators are determined as eigenvalues
$\Delta_\pm$ of the matrix $\Delta^{\Op{}} =
d^{\Op{}}+\gamma^{\Op{}}(g_1^*,g_2^*)$, where $d^{\Op{}}$ is a matrix of the
canonical dimensions of the operators, and $\gamma^{\Op{}}(g_1^*,g_2^*)$ are
anomalous dimensions from equation~\eqref{eq:gammaO} at the fixed point. In our
case, the canonical dimensions of the operators are equal and proportional to the unit matrix
$d^{\Op{}} = (d-2) \cdot \mathbb{1}$. Dimensions of these operators can be
identified with critical exponents in the
$O(n)$-universality class. While the definition of the anomalous dimensions
depends on the
particular model (non-linear sigma model, $O(n)$-symmetric $\vphi^4$ model, or
$O(n)$-symmetric $\vphi^3$ model), the critical dimensions of operators can
easily be
mapped from one model to another. One can show that $\Delta_\pm$ must
coincide with $\Delta^{\mathrm{NLSM}}_{\psi^2}$ and
$\Delta^{\mathrm{NLSM}}_{\phi^2}$ from the non-linear sigma
model~\cite{Vasilev}. This allow us to relate these dimensions to critical exponents of the
$O(n)$-universality class, and to compare with results of the $1/n$ expansion,
\begin{equation}
  \label{eq:largeNdelta}
  \Delta_{+} = \Delta^{\mathrm{NLSM}}_{\psi^2} = d + \omega,
  \quad
  \Delta_{-} = \Delta^{\mathrm{NLSM}}_{\phi^2} = d - \Delta_{\sigma} = \frac{1}{\nu}.
\end{equation}
Here critical dimension $\Delta_{+}$ agrees with the correction-to-scaling exponent $\omega$ calculated to order $1/n^2$
in Refs.~\cite{Broadhurst:1996ur,Gracey:1996ub}, while the critical dimension
$\Delta_{-}$ is the already calculated exponent $1/\nu$.
Another question we can address with our results for the beta
functions~\eqref{eq:betaDefPoles} is about $n_{\rm cr}$: the critical value of
$n$ below which the fixed point becomes complex. A very large value $n_{\rm
cr}=1038$ was determined at leading order for the first time in
Ref.~\cite{Fei:2014yja} and refined using higher-loop results for the
beta-functions in~\cite{Fei:2014xta,Gracey:2015tta}. It was further considered
in~\cite{Eichhorn:2016hdi} with functional renormalization-group methods. Let us
follow Ref.~\cite{Fei:2014yja}, where the following conditions determine $n_{\rm
cr}$,
\begin{equation}
  \label{eq:ncrcond}
  \beta_1 = 0,
  \quad
  \beta_2 = 0,
  \quad
  \frac{\partial \beta_1}{\partial g_1}\frac{\partial \beta_2}{\partial g_2} -
  \frac{\partial \beta_1}{\partial g_2}\frac{\partial \beta_2}{\partial g_1} = 0.
\end{equation}
Using an ansatz for $n_{\rm cr}$ as a series in $\ep$, and rescaled coupling
constants $(x,y)$, defined as $g_1=2\sqrt{\frac{3\ep}{n}}x,
g_2=2\sqrt{\frac{3\ep}{n}}y$, we determine a set of critical values,
\begin{align}
  \label{eq:Ncrit5l}
  n_{\rm cr} & = 
               1038.2661
               -1219.6796 \ep
               -1456.6933 \ep^2
               + 3621.6847 \ep^3
               + 986.2232 \ep^4
               + O\left(\ep^5\right)\nonumber\\
  x_{\rm cr} & =
               1.01804
               - 0.0187953 \ep
               + 0.0276061 \ep^2
               - 0.0258653 \ep^3
               + 0.188983 \ep^4
               + O\left(\ep^5\right)\nonumber\\
  y_{\rm cr} & =
               8.90305
               - 0.420480 \ep
               + 4.06718 \ep^2
               - 2.00946 \ep^3
               - 62.4036 \ep^4
               + O\left(\ep^5\right)
\end{align}
From the multiple solutions of the system~\eqref{eq:ncrcond}, we have choosen
the one corresponding to the large-$n$ limit of the theory.
To get an estimate for $n_{\rm cr}$ we use Pade approximants. It is well known that
for finite values of the expansion parameter estimates obtained from different
approximants may differ significantly. This happens when approximants
have spurious poles or a powerlaw asymptotics becomes dominant. It is hard to
estimate the influence of these factors, and the choice of a particular
approximant becomes intricate, and biased by already known estimates from
other methods. To avoid this, we use the method suggested in
Ref.~\cite{Adzhemyan:2019gvv}, excluding approximants for which the influence of poles or
asymptotics is obvious, and consider estimates from the remaining approximants as
``independent measurements''. This allows us to get unbiased estimate as well as
error bars. As a result, $n_{\rm cr}=340\pm200$ at four loops and $n_{\rm
cr}=420\pm120$ at five loops, both in dimension $d=5$. These values are in agreement with
earlier estimates~\cite{Fei:2014yja,Fei:2014xta,Gracey:2015tta}.
\section{Conclusion}
\label{sec:conclusion}
We calculated the renormalization-group functions and the corresponding critical
exponents for the simplest scalar cubic theory and its $O(n)$-extension in
$d=6-2\ep$ space-time dimensions at five-loop order. To achieve this, we
constructed an $\ep$-finite basis of fully massive five-loop tadpole integrals,
sufficient for computation of all needed renormalization constants. For each
diagram in the scalar theory, we provide counterterms, which will help extend
our results to more advanced field theory models and obtain the critical
dimensions of other observables of interest. The critical exponents in the
$O(n)$-symmetric model were checked to agree with results of the $1/n$-expansion
available in the literature.
\section*{Acknowledgements}
\label{sec:acknowledgments}
We thank A.Bednyakov, G.Kalagov, N.Lebedev and K.Wiese for fruitfull discussions
and careful reading of the manuscript, J.Gracey for his comments on the
large-$n$ limit of the $O(n)$ model, as well as M.Borinsky and O.Schnetz for
sharing their results prior to publication. We are grateful to the Joint
Institute for Nuclear Research for letting us use their
supercomputer``Govorun''. This work is supported by the Foundation for the
Advancement of Theoretical Physics and Mathematics ``BASIS.''

\appendix

\section{Anomalous dimensions for scalar theory}
\label{sec:AnDimPhi3}

\begin{align}
  \gamma_{\vphi} = & \frac{1}{12}g^2 + \frac{13 }{432}g^4
                     + \left(\frac{5195}{62208} - \frac{\zeta_3}{24}\right) g^6
                     +  \left(\frac{53449}{248832} + \frac{35 \zeta_3}{864} + \frac{7 \zeta_4}{96}
                     - \frac{5 \zeta_5}{18}\right) g^8 + \nonumber\\
                   & \left(\frac{16492987}{20155392} + \frac{56693 \zeta_3}{62208} 
                     + \frac{5651 \zeta_4}{27648} - \frac{4471 \zeta_5}{10368} + \frac{25 \zeta_3^2}{144}
                     + \frac{125 \zeta_6}{288} - \frac{147 \zeta_7}{64}\right) g^{10}  + O\left(g^{11}\right)
                     \label{eq:gamPhiPhi3}\\
  \gamma_{m^2} =  & \frac{5}{6} g^2 + \frac{97}{108} g^4
                    + \left(\frac{52225}{31104} + \frac{7 \zeta_3}{12}\right) g^6
                    + \left(\frac{445589}{93312} + \frac{821 \zeta_3}{144} - \frac{19 \zeta_4}{48}
                    - \frac{35 \zeta_5}{18}\right) g^8 + \nonumber\\
                   & \left(\frac{40331135}{2519424} + \frac{839129 \zeta_3}{15552} 
                     - \frac{66953 \zeta_4}{13824} + \frac{225457 \zeta_5}{5184} + \frac{229 \zeta_3^2}{36}
                     + \frac{25 \zeta_6}{9} - \frac{2821 \zeta_7}{32}\right) g^{10} + O\left(g^{11}\right)
                     \label{eq:gamMMPhi3}
\end{align}

\section{Results for the $O(n)$ symmetric model}
\label{sec:resOn}

\begin{align}
  g_1^* & = 2\sqrt{\frac{3 \ep}{n}}\left[1+
          \left(\frac{22}{n}+\frac{726}{n^2}-\frac{326180}{n^3} + \mathcal{O}\left(\frac{1}{n^4}\right)\right)\right.\nonumber\\
        & +\left(-\frac{155}{3 n} - \frac{3410}{n^2}
          + \frac{1825090}{n^3}
          + \mathcal{O}\left(\frac{1}{n^4}\right) \right)\ep \nonumber\\
        & + \left( \frac{1777}{36 n}
          + \left(\frac{29093}{9} - 4680 \zeta_3\right)\frac{1}{n^2}
          + \left(-\frac{106755739}{18} - 912240 \zeta_3\right)\frac{1}{n^3}
          + \mathcal{O}\left(\frac{1}{n^4}\right)\right)\ep^2\nonumber\\
        & + \left(-\frac{217}{324 n} + \left(\frac{709151}{108} + 25008 \zeta_3 - 4680 \zeta_4 + 2880 \zeta_5\right)\frac{1}{n^2}\right.\nonumber\\
        & + \left.\left( \frac{779869165}{54} + 17839320 \zeta_3 - 912240 \zeta_4 - 19509120 \zeta_5\right)\frac{1}{n^3}
          + \mathcal{O}\left(\frac{1}{n^4}\right)\right)\ep^3 \nonumber\\
        & + \left(\frac{6973}{2592} - \frac{155 \zeta_3}{6}\right)\frac{1}{n}
          + \left(-\frac{49050023}{7776} - \frac{81742 \zeta_3}{3} + 
          1296 \zeta_3^2 + 28134 \zeta_4 - 69120 \zeta_5 + 
          5400 \zeta_6\right)\frac{1}{n^2}\nonumber\\
        & + \left(-\frac{17059272503}{972} - \frac{77297317 \zeta_3}{3} - 
          5870880 \zeta_3^2 + 20039010 \zeta_4 + 92059740 \zeta_5 \right.\nonumber\\
        & \left.\left.- 36579600 \zeta_6 - 
          23528232 \zeta_7\Bigr)\frac{1}{n^3} + \mathcal{O}\left(\frac{1}{n^4}\right) \right)\ep^4 + O(\ep^5)\right]
          \label{eq:g1crit}\\
  g_2^* & = 2\sqrt{\frac{3 \ep}{n}}\left[6 +
          \left(\frac{972}{n} + \frac{412596}{n^2} + \frac{247346520}{n^3} + \mathcal{O}\left(\frac{1}{n^4}\right)\right)\right.\nonumber\\
        & + \left(-\frac{1290}{n} - \frac{1036020}{n^2} - \frac{908667180}{n^3} + \mathcal{O}\left(\frac{1}{n^4}\right)\right)\ep\nonumber\\
        & + \left(\frac{2781}{2 n} + \left(1083644 - 628560 \zeta_3\right)\frac{1}{n^2}
          + \left(1156981601 - 632020320 \zeta_3\right)\frac{1}{n^3} + \mathcal{O}\left(\frac{1}{n^4}\right)\right)\ep^2\nonumber\\
        & + \left(-\frac{3461}{18 n}
          + \left( \frac{25209239}{18} + 5561280 \zeta_3 - 628560 \zeta_4 - 3369600 \zeta_5\right)\frac{1}{n^2}\right.\nonumber\\
        & + \left.\left(\frac{18166643735}{9} + 9509192880 \zeta_3 - 758121120 \zeta_4 - 8137843200 \zeta_5\right)\frac{1}{n^3} + \mathcal{O}\left(\frac{1}{n^4}\right)\right)\ep^3\nonumber\\
        & + \left(\left(\frac{5945}{144} - 645 \zeta_3\right)\frac{1}{n}
          + \left(-\frac{1204206197}{432} - 11677092 \zeta_3 - 1516320 \zeta_3^2 + 6256440 \zeta_4  + 10069920 \zeta_5\right.\right.
          \nonumber\\
        & - 6318000 \zeta_6\Bigr)\frac{1}{n^2} - \left(
          \frac{143586895124}{27} + 11476374450 \zeta_3 + 2300337792 \zeta_3^2 - 11687633640 \zeta_4\right.\nonumber\\
        & \left.\left.\left.  - 48410098200 \zeta_5 + 16147512000 \zeta_6 + 30589940304 \zeta_7
          \right)\frac{1}{n^3} + \mathcal{O}\left(\frac{1}{n^4}\right)\right)\ep^4 + \mathcal{O}(\ep^5)\right]
          \label{eq:g2crit}
\end{align}

\section{Diagrams calculated manually}
\label{sec:manual-dias}
Here we present details of the $\KRp$ operation applied to six integrals with
subdivergencies from the set $J^{(3)}$. Since $\KRp$ is independent of
external momenta and masses, we made use of results for $\KRp$ for
logarithmically divergent massive tadpole integrals from previous steps. Thin
black lines are massless propagators, while thick red ones are massive.
\tikzset{ tad/.style={line width=1pt,draw=red}}
\begin{alignat}{2}
  \KRp G_{130} &= 
  \KRp
  \left[
    \vcenter{\hbox{
        \begin{tikzpicture}[use Hobby shortcut, scale=0.8]
          \draw (0,0) circle (1);
          \coordinate (v0) at (0:1);
          \coordinate (v30) at (30:1);
          \coordinate (v45) at (45:1);
          \coordinate (v60) at (60:1);   
          \coordinate (v90) at (90:1);
          \coordinate (v120) at (120:1);
          \coordinate (v135) at (135:1);
          \coordinate (v150) at (150:1);
          \coordinate (v180) at (180:1);
          \coordinate (v210) at (210:1);
          \coordinate (v225) at (225:1);
          \coordinate (v240) at (240:1);
          \coordinate (v270) at (270:1);
          \coordinate (v300) at (300:1);
          \coordinate (v315) at (315:1);
          \coordinate (v330) at (330:1);
          \coordinate (u1) at ($(v180)!0.3!(v0)+(0,-0.3)$);
          \coordinate (u2) at ($(v0)!0.3!(v180)+(0,-0.3)$);
          \draw (v180) .. (u1) .. (u2) .. (v0);
          \draw (u1) -- (v240);
          \draw (u2) -- (v300);
          \draw (v135) .. (90:0.5) .. (v45);
          \draw (v210) -- (210:1.2);
          \draw (v330) -- (330:1.2);
          \draw (v90) -- (90:1.2);
          \fill (v180) circle (1pt);
          \fill (v0) circle (1pt);
          \fill (u1) circle (1pt);
          \fill (u2) circle (1pt);
          \fill (v90) circle (1pt);
          \fill (v210) circle (1pt);
          \fill (v330) circle (1pt);
          \fill (v240) circle (1pt);
          \fill (v300) circle (1pt);
          \fill (v45) circle (1pt);
          \fill (v135) circle (1pt);
        \end{tikzpicture}
      }}
  \right]
  && =
  \mathcal{K}\left[G(1,2+4\ep)
    \vcenter{\hbox{
        \begin{tikzpicture}[use Hobby shortcut, scale=0.8]
          \draw (0,0) circle (0.7);
          \coordinate (v0) at (0:0.7);
          \coordinate (v30) at (30:0.7);
          \coordinate (v45) at (45:0.7);
          \coordinate (v60) at (60:0.7);   
          \coordinate (v90) at (90:0.7);
          \coordinate (v120) at (120:0.7);
          \coordinate (v135) at (135:0.7);
          \coordinate (v150) at (150:0.7);
          \coordinate (v180) at (180:0.7);
          \coordinate (v210) at (210:0.7);
          \coordinate (v225) at (225:0.7);
          \coordinate (v240) at (240:0.7);
          \coordinate (v270) at (270:0.7);
          \coordinate (v300) at (300:0.7);
          \coordinate (v315) at (315:0.7);
          \coordinate (v330) at (330:0.7);
          \draw (v135) -- (v225);
          \draw (v45) -- (v315);
          \draw (v120) .. (90:0.5) .. (v60);
          \draw (v180) -- (180:1);
          \draw (v0) -- (0:1);
          \fill (v180) circle (1pt);
          \fill (v0) circle (1pt);
          \fill (v45) circle (1pt);
          \fill (v135) circle (1pt);
          \fill (v225) circle (1pt);
          \fill (v315) circle (1pt);
          \fill (v60) circle (1pt);
          \fill (v120) circle (1pt);
          \fill (v90) circle (1pt);
          \fill (155:0.7) circle (1pt);
          \fill (25:0.7) circle (1pt);
        \end{tikzpicture}
      }}
  \right]
  - \mathcal{K}\left[ \mathcal{K}\left[
      \vcenter{\hbox{
          \begin{tikzpicture}[use Hobby shortcut, scale=0.8]
            \draw[tad] (0,0) circle (0.5);
            \fill (90:0.5) circle (1pt);
            \fill (210:0.5) circle (1pt);
            \fill (330:0.5) circle (1pt);
          \end{tikzpicture}
        }}
    \right]
    \vcenter{\hbox{
        \begin{tikzpicture}[use Hobby shortcut, scale=0.8]
          \draw (0,0) circle (0.7);
          \coordinate (v0) at (0:0.7);
          \coordinate (v30) at (30:0.7);
          \coordinate (v45) at (45:0.7);
          \coordinate (v60) at (60:0.7);   
          \coordinate (v90) at (90:0.7);
          \coordinate (v120) at (120:0.7);
          \coordinate (v135) at (135:0.7);
          \coordinate (v150) at (150:0.7);
          \coordinate (v180) at (180:0.7);
          \coordinate (v210) at (210:0.7);
          \coordinate (v225) at (225:0.7);
          \coordinate (v240) at (240:0.7);
          \coordinate (v270) at (270:0.7);
          \coordinate (v300) at (300:0.7);
          \coordinate (v315) at (315:0.7);
          \coordinate (v330) at (330:0.7);
          \coordinate (u1) at (-0.3,0);
          \coordinate (u2) at (0.3,0);
          \draw (v120) -- (u1);
          \draw (v180) -- (u1);
          \draw (v60) -- (u2);
          \draw (v0) -- (u2);
          \draw (u1) -- (u2);
          \draw (v180) -- (180:1);
          \draw (v0) -- (0:1);
          \fill (v180) circle (1pt);
          \fill (v0) circle (1pt);
          \fill (v30) circle (1pt);
          \fill (v60) circle (1pt);
          \fill (v90) circle (1pt);
          \fill (v120) circle (1pt);
          \fill (v150) circle (1pt);
          \fill (u1) circle (1pt);
          \fill (u2) circle (1pt);
        \end{tikzpicture}
      }}
  \right]\label{eq:KRg130}\\
  \KRp G_{848} & = \KRp
  \left[
    \vcenter{\hbox{
        \begin{tikzpicture}[use Hobby shortcut, scale=0.8]
          \draw (0,0) circle (1);
          \coordinate (v0) at (0:1);
          \coordinate (v30) at (30:1);
          \coordinate (v45) at (45:1);
          \coordinate (v60) at (60:1);   
          \coordinate (v90) at (90:1);
          \coordinate (v120) at (120:1);
          \coordinate (v135) at (135:1);
          \coordinate (v150) at (150:1);
          \coordinate (v180) at (180:1);
          \coordinate (v190) at (190:1);
          \coordinate (v210) at (210:1);
          \coordinate (v225) at (225:1);
          \coordinate (v240) at (240:1);
          \coordinate (v270) at (270:1);
          \coordinate (v300) at (300:1);
          \coordinate (v315) at (315:1);
          \coordinate (v330) at (330:1);
          \coordinate (v350) at (350:1);
          \coordinate (u1) at ($(v135)!0.3!(v45)+(0,-0.1)$);
          \coordinate (u2) at ($(v45)!0.3!(v135)+(0,-0.1)$);
          \coordinate (u3) at ($(v190)!0.3!(v350)$);
          \coordinate (u4) at ($(v350)!0.3!(v190)$);
          \draw (v45) .. (u2) ..(u1) .. (v135);
          \draw (v190) -- (v350);
          \draw (u1) -- (u3);
          \draw (u2) -- (u4);
          \draw (v210) -- (210:1.2);
          \draw (v330) -- (330:1.2);
          \draw (v90) -- (90:1.2);
          \fill (u1) circle (1pt);
          \fill (u2) circle (1pt);
          \fill (u3) circle (1pt);
          \fill (u4) circle (1pt);
          \fill (v90) circle (1pt);
          \fill (v190) circle (1pt);
          \fill (v350) circle (1pt);
          \fill (v45) circle (1pt);
          \fill (v135) circle (1pt);
          \fill (v210) circle (1pt);
          \fill (v330) circle (1pt);
        \end{tikzpicture}
      }}
  \right]
  && =
  \mathcal{K}\left[G(1,2+4\ep)
    \vcenter{\hbox{
        \begin{tikzpicture}[use Hobby shortcut, scale=0.8]
          \draw (0,0) circle (0.7);
          \coordinate (v0) at (0:0.7);
          \coordinate (v30) at (30:0.7);
          \coordinate (v45) at (45:0.7);
          \coordinate (v60) at (60:0.7);   
          \coordinate (v90) at (90:0.7);
          \coordinate (v120) at (120:0.7);
          \coordinate (v135) at (135:0.7);
          \coordinate (v150) at (150:0.7);
          \coordinate (v180) at (180:0.7);
          \coordinate (v210) at (210:0.7);
          \coordinate (v225) at (225:0.7);
          \coordinate (v240) at (240:0.7);
          \coordinate (v270) at (270:0.7);
          \coordinate (v300) at (300:0.7);
          \coordinate (v315) at (315:0.7);
          \coordinate (v330) at (330:0.7);
          \coordinate (u1) at (-0.3,0);
          \coordinate (u2) at (0.3,0);
          \draw (v135) -- (u1);
          \draw (v225) -- (u1);
          \draw (v45) -- (u2);
          \draw (v315) -- (u2);
          \draw (u1) -- (u2);
          \draw (v180) -- (180:1);
          \draw (v0) -- (0:1);
          \fill (v180) circle (1pt);
          \fill (v0) circle (1pt);
          \fill (v45) circle (1pt);
          \fill (v135) circle (1pt);
          \fill (v225) circle (1pt);
          \fill (v315) circle (1pt);
          \fill (v90) circle (1pt);
          \fill (u1) circle (1pt);
          \fill (u2) circle (1pt);
        \end{tikzpicture}
      }}
  \right]
  - \mathcal{K}\left[\mathcal{K}\left[
      \vcenter{\hbox{
          \begin{tikzpicture}[use Hobby shortcut, scale=0.8]
            \draw[tad] (0,0) circle (0.7);
            \coordinate (v0) at (0:0.7);
            \coordinate (v30) at (30:0.7);
            \coordinate (v45) at (45:0.7);
            \coordinate (v60) at (60:0.7);   
            \coordinate (v90) at (90:0.7);
            \coordinate (v120) at (120:0.7);
            \coordinate (v135) at (135:0.7);
            \coordinate (v150) at (150:0.7);
            \coordinate (v180) at (180:0.7);
            \coordinate (v210) at (210:0.7);
            \coordinate (v225) at (225:0.7);
            \coordinate (v240) at (240:0.7);
            \coordinate (v270) at (270:0.7);
            \coordinate (v300) at (300:0.7);
            \coordinate (v315) at (315:0.7);
            \coordinate (v330) at (330:0.7);
            \coordinate (u1) at (-0.3,0);
            \coordinate (u2) at (0.3,0);
            \draw[tad] (v135) -- (u1);
            \draw[tad] (v225) -- (u1);
            \draw[tad] (v45) -- (u2);
            \draw[tad] (v315) -- (u2);
            \draw[tad] (u1) -- (u2);
            \fill (v180) circle (1pt);
            \fill (v0) circle (1pt);
            \fill (v45) circle (1pt);
            \fill (v135) circle (1pt);
            \fill (v225) circle (1pt);
            \fill (v315) circle (1pt);
            \fill (v90) circle (1pt);
            \fill (u1) circle (1pt);
            \fill (u2) circle (1pt);
          \end{tikzpicture}
        }}
    \right]
    \vcenter{\hbox{
        \begin{tikzpicture}[use Hobby shortcut, scale=0.8]
          \draw (0,0) circle (0.5);
          \draw (180:0.5) -- (180:0.7);
          \draw (0:0.5) -- (0:0.7);
          \fill (180:0.5) circle (1pt);
          \fill (0:0.5) circle (1pt);
          \fill (90:0.5) circle (1pt);
        \end{tikzpicture}
      }}
  \right]
  \label{eq:KRg848}\\
  \KRp G_{2} & = \KRp\left[
    \vcenter{\hbox{
        \begin{tikzpicture}[use Hobby shortcut, scale=0.8]
          \draw (0,0) circle (1);
          \coordinate (v0) at (0:1);
          \coordinate (v30) at (30:1);
          \coordinate (v45) at (45:1);
          \coordinate (v60) at (60:1);   
          \coordinate (v90) at (90:1);
          \coordinate (v120) at (120:1);
          \coordinate (v135) at (135:1);
          \coordinate (v150) at (150:1);
          \coordinate (v180) at (180:1);
          \coordinate (v190) at (190:1);
          \coordinate (v210) at (210:1);
          \coordinate (v225) at (225:1);
          \coordinate (v240) at (240:1);
          \coordinate (v270) at (270:1);
          \coordinate (v300) at (300:1);
          \coordinate (v315) at (315:1);
          \coordinate (v330) at (330:1);
          \coordinate (v350) at (350:1);
          \coordinate (u1) at (30:0.7);
          \coordinate (u2) at (150:0.7);
          \coordinate (c1) at ($(v315)!0.5!(u2)$);
          \coordinate (c2) at ($(u2)!0.4!(v315)$);
          \draw (v60) .. (u1) .. (v0);
          \draw (v120) .. (u2) .. (v180);
          \draw (v225) -- (u1);
          \draw (v315) -- (c1);
          \draw (u2) -- (c2);
          \draw (v210) -- (210:1.2);
          \draw (v330) -- (330:1.2);
          \draw (v90) -- (90:1.2);
          \fill (u1) circle (1pt);
          \fill (u2) circle (1pt);
          \fill (v90) circle (1pt);
          \fill (v60) circle (1pt);
          \fill (v120) circle (1pt);
          \fill (v0) circle (1pt);
          \fill (v180) circle (1pt);
          \fill (v225) circle (1pt);
          \fill (v315) circle (1pt);
          \fill (v210) circle (1pt);
          \fill (v330) circle (1pt);
        \end{tikzpicture}
      }}
  \right]
  && =
  \mathcal{K}\left[G(1,2+4\ep)
    \vcenter{\hbox{
        \begin{tikzpicture}[use Hobby shortcut, scale=0.8]
          \draw (0,0) circle (0.7);
          \coordinate (v0) at (0:0.7);
          \coordinate (v30) at (30:0.7);
          \coordinate (v45) at (45:0.7);
          \coordinate (v60) at (60:0.7);   
          \coordinate (v90) at (90:0.7);
          \coordinate (v120) at (120:0.7);
          \coordinate (v135) at (135:0.7);
          \coordinate (v150) at (150:0.7);
          \coordinate (v180) at (180:0.7);
          \coordinate (v210) at (210:0.7);
          \coordinate (v225) at (225:0.7);
          \coordinate (v240) at (240:0.7);
          \coordinate (v270) at (270:0.7);
          \coordinate (v300) at (300:0.7);
          \coordinate (v315) at (315:0.7);
          \coordinate (v330) at (330:0.7);
          \coordinate (u1) at (0,-0.3);
          \coordinate (u2) at (0,0.3);
          \draw (v135) -- (u2);
          \draw (v225) -- (u1);
          \draw (v45) -- (u2);
          \draw (v315) -- (u1);
          \draw (u1) -- (u2);
          \draw (v180) -- (180:1);
          \draw (v0) -- (0:1);
          \fill (v180) circle (1pt);
          \fill (v0) circle (1pt);
          \fill (v45) circle (1pt);
          \fill (v135) circle (1pt);
          \fill (v225) circle (1pt);
          \fill (v315) circle (1pt);
          \fill (155:0.7) circle (1pt);
          \fill (-25:0.7) circle (1pt);
          \fill (0,0) circle (1pt);
          \fill (u1) circle (1pt);
          \fill (u2) circle (1pt);
        \end{tikzpicture}
      }}
  \right]
  - 2 \mathcal{K}\left[ \mathcal{K}\left[
      \vcenter{\hbox{
          \begin{tikzpicture}[use Hobby shortcut, scale=0.8]
            \draw[tad] (0,0) circle (0.5);
            \fill (90:0.5) circle (1pt);
            \fill (210:0.5) circle (1pt);
            \fill (330:0.5) circle (1pt);
          \end{tikzpicture}
        }}
    \right]
    \vcenter{\hbox{
        \begin{tikzpicture}[use Hobby shortcut, scale=0.8]
          \draw (0,0) circle (0.7);
          \coordinate (v0) at (0:0.7);
          \coordinate (v30) at (30:0.7);
          \coordinate (v45) at (45:0.7);
          \coordinate (v60) at (60:0.7);   
          \coordinate (v90) at (90:0.7);
          \coordinate (v120) at (120:0.7);
          \coordinate (v135) at (135:0.7);
          \coordinate (v150) at (150:0.7);
          \coordinate (v165) at (165:0.7);
          \coordinate (v180) at (180:0.7);
          \coordinate (v210) at (210:0.7);
          \coordinate (v225) at (225:0.7);
          \coordinate (v240) at (240:0.7);
          \coordinate (v270) at (270:0.7);
          \coordinate (v300) at (300:0.7);
          \coordinate (v315) at (315:0.7);
          \coordinate (v330) at (330:0.7);
          \coordinate (u1) at (135:0.85);
          \coordinate (c1) at ($(v45)!0.35!(v180)$);
          \coordinate (c2) at ($(v180)!0.50!(v45)$);
          \draw (v150) .. (u1) .. (v120);
          \draw (v0) -- (v135);
          \draw (v45) -- (c1);
          \draw (v180) -- (c2);
          \draw (v180) -- (180:1);
          \draw (v0) -- (0:1);
          \fill (v180) circle (1pt);
          \fill (v0) circle (1pt);
          \fill (v165) circle (1pt);
          \fill (v120) circle (1pt);
          \fill (v90) circle (1pt);
          \fill (v150) circle (1pt);
          \fill (v135) circle (1pt);
          \fill (25:0.7) circle (1pt);
          \fill (v45) circle (1pt);
        \end{tikzpicture}
      }}
  \right]
  \nonumber
  \\
  & && +  \mathcal{K}\left[ \mathcal{K}\left[
      \vcenter{\hbox{
          \begin{tikzpicture}[use Hobby shortcut, scale=0.8]
            \draw[tad] (0,0) circle (0.5);
            \fill (90:0.5) circle (1pt);
            \fill (210:0.5) circle (1pt);
            \fill (330:0.5) circle (1pt);
          \end{tikzpicture}
        }}
    \right]^2
    \vcenter{\hbox{
        \begin{tikzpicture}[use Hobby shortcut, scale=0.8]
          \draw (0,0) circle (0.7);
          \coordinate (v0) at (0:0.7);
          \coordinate (v30) at (30:0.7);
          \coordinate (v45) at (45:0.7);
          \coordinate (v60) at (60:0.7);   
          \coordinate (v90) at (90:0.7);
          \coordinate (v120) at (120:0.7);
          \coordinate (v135) at (135:0.7);
          \coordinate (v150) at (150:0.7);
          \coordinate (v165) at (165:0.7);
          \coordinate (v180) at (180:0.7);
          \coordinate (v210) at (210:0.7);
          \coordinate (v225) at (225:0.7);
          \coordinate (v240) at (240:0.7);
          \coordinate (v270) at (270:0.7);
          \coordinate (v300) at (300:0.7);
          \coordinate (v315) at (315:0.7);
          \coordinate (v330) at (330:0.7);
          \coordinate (u1) at (135:0.85);
          \coordinate (c1) at ($(v45)!0.35!(v180)$);
          \coordinate (c2) at ($(v180)!0.50!(v45)$);
          \draw (v0) -- (v135);
          \draw (v45) -- (c1);
          \draw (v180) -- (c2);
          \draw (v180) -- (180:1);
          \draw (v0) -- (0:1);
          \fill (v180) circle (1pt);
          \fill (v0) circle (1pt);
          \fill (v90) circle (1pt);
          \fill (v135) circle (1pt);
          \fill (25:0.7) circle (1pt);
          \fill (155:0.7) circle (1pt);
          \fill (v45) circle (1pt);
        \end{tikzpicture}
      }}
  \right]
  \label{eq:KRg2}\\
  \KRp G_{916} & = \KRp\left[
    \vcenter{\hbox{
        \begin{tikzpicture}[use Hobby shortcut, scale=0.8]
          \draw (0,0) circle (1);
          \coordinate (v0) at (0:1);
          \coordinate (v10) at (10:1);
          \coordinate (v30) at (30:1);
          \coordinate (v45) at (45:1);
          \coordinate (v60) at (60:1);   
          \coordinate (v90) at (90:1);
          \coordinate (v120) at (120:1);
          \coordinate (v135) at (135:1);
          \coordinate (v150) at (150:1);
          \coordinate (v170) at (170:1);
          \coordinate (v180) at (180:1);
          \coordinate (v190) at (190:1);
          \coordinate (v210) at (210:1);
          \coordinate (v225) at (225:1);
          \coordinate (v240) at (240:1);
          \coordinate (v270) at (270:1);
          \coordinate (v300) at (300:1);
          \coordinate (v315) at (315:1);
          \coordinate (v330) at (330:1);
          \coordinate (v350) at (350:1);
          \coordinate (u10) at ($(v180)!0.25!(v0)$);
          \coordinate (u20) at ($(v0)!0.25!(v180)$);
          \coordinate (u1) at (u10|-1,-0);
          \coordinate (u2) at (u20|-1,-0);
          \coordinate (c1) at ($(u2)!0.32!(v135)$);
          \coordinate (c2) at ($(v135)!0.53!(u2)$);
          \draw (v170) .. (u1) .. (u2) .. (v10);
          \draw (v190) .. (0,-0.4) .. (v350);
          \draw (u1) -- (v45);
          \draw (u2) -- (c1);
          \draw (v135) -- (c2);
          \draw (v210) -- (210:1.2);
          \draw (v330) -- (330:1.2);
          \draw (v90) -- (90:1.2);
          \fill (u1) circle (1pt);
          \fill (u2) circle (1pt);
          \fill (v90) circle (1pt);
          \fill (v190) circle (1pt);
          \fill (v350) circle (1pt);
          \fill (v45) circle (1pt);
          \fill (v135) circle (1pt);
          \fill (v210) circle (1pt);
          \fill (v330) circle (1pt);
          \fill (v10) circle (1pt);
          \fill (v170) circle (1pt);
        \end{tikzpicture}
      }}
  \right]
  && =
  \mathcal{K}\left[
    G(1,2+4\ep)
    \vcenter{\hbox{
        \begin{tikzpicture}[use Hobby shortcut, scale=0.8]
          \draw (0,0) circle (0.7);
          \coordinate (v0) at (0:0.7);
          \coordinate (v30) at (30:0.7);
          \coordinate (v45) at (45:0.7);
          \coordinate (v60) at (60:0.7);   
          \coordinate (v90) at (90:0.7);
          \coordinate (v120) at (120:0.7);
          \coordinate (v135) at (135:0.7);
          \coordinate (v150) at (150:0.7);
          \coordinate (v180) at (180:0.7);
          \coordinate (v210) at (210:0.7);
          \coordinate (v225) at (225:0.7);
          \coordinate (v240) at (240:0.7);
          \coordinate (v270) at (270:0.7);
          \coordinate (v300) at (300:0.7);
          \coordinate (v315) at (315:0.7);
          \coordinate (v330) at (330:0.7);
          \coordinate (u1) at (-0.3,0);
          \coordinate (u2) at (0.3,0);
          \coordinate (c1) at ($(u1)!0.40!(v60)$);
          \coordinate (c2) at ($(v60)!0.45!(u1)$);
          \draw (160:0.7) .. (u1) .. (u2) .. (20:0.7);
          \draw (u2) -- (v120);
          \draw (u1) -- (c1);
          \draw (v60) -- (c2);
          \draw (v180) -- (180:1);
          \draw (v0) -- (0:1);
          \fill (v180) circle (1pt);
          \fill (v0) circle (1pt);
          \fill (v60) circle (1pt);
          \fill (v120) circle (1pt);
          \fill (v90) circle (1pt);
          \fill (u1) circle (1pt);
          \fill (u2) circle (1pt);
          \fill (160:0.7) circle (1pt);
          \fill (20:0.7) circle (1pt);
        \end{tikzpicture}
      }}
  \right]
  - \mathcal{K}\left[\KRp\left[
      \vcenter{\hbox{
          \begin{tikzpicture}[use Hobby shortcut, scale=0.8]
            \draw[tad] (0,0) circle (0.7);
            \coordinate (v0) at (0:0.7);
            \coordinate (v30) at (30:0.7);
            \coordinate (v45) at (45:0.7);
            \coordinate (v60) at (60:0.7);   
            \coordinate (v90) at (90:0.7);
            \coordinate (v120) at (120:0.7);
            \coordinate (v135) at (135:0.7);
            \coordinate (v150) at (150:0.7);
            \coordinate (v180) at (180:0.7);
            \coordinate (v210) at (210:0.7);
            \coordinate (v225) at (225:0.7);
            \coordinate (v240) at (240:0.7);
            \coordinate (v270) at (270:0.7);
            \coordinate (v300) at (300:0.7);
            \coordinate (v315) at (315:0.7);
            \coordinate (v330) at (330:0.7);
            \coordinate (u1) at (-0.3,-0.15);
            \coordinate (u2) at (0.3,-0.15);
            \coordinate (c1) at ($(v45)!0.55!(u1)$);
            \coordinate (c2) at ($(u1)!0.3!(v45)$);
            \draw[tad] (v180) .. (u1) .. (u2) .. (v0);
            \draw[tad] (v135) -- (u2);
            \draw[tad] (v45) -- (c1);
            \draw[tad] (u1) -- (c2);
            \fill (v180) circle (1pt);
            \fill (v0) circle (1pt);
            \fill (v45) circle (1pt);
            \fill (v135) circle (1pt);
            \fill (v225) circle (1pt);
            \fill (v315) circle (1pt);
            \fill (v90) circle (1pt);
            \fill (u1) circle (1pt);
            \fill (u2) circle (1pt);
          \end{tikzpicture}
        }}
    \right]
    \vcenter{\hbox{
        \begin{tikzpicture}[use Hobby shortcut, scale=0.8]
          \draw (0,0) circle (0.5);
          \draw (180:0.5) -- (180:0.7);
          \draw (0:0.5) -- (0:0.7);
          \fill (180:0.5) circle (1pt);
          \fill (0:0.5) circle (1pt);
          \fill (90:0.5) circle (1pt);
        \end{tikzpicture}
      }}
  \right]
  \nonumber
  \\
  &    && - \mathcal{K}\left[\mathcal{K}\left[
      \vcenter{\hbox{
          \begin{tikzpicture}[use Hobby shortcut, scale=0.8]
            \draw[tad] (0,0) circle (0.7);
            \coordinate (v0) at (0:0.7);
            \coordinate (v30) at (30:0.7);
            \coordinate (v45) at (45:0.7);
            \coordinate (v60) at (60:0.7);   
            \coordinate (v90) at (90:0.7);
            \coordinate (v120) at (120:0.7);
            \coordinate (v135) at (135:0.7);
            \coordinate (v150) at (150:0.7);
            \coordinate (v180) at (180:0.7);
            \coordinate (v210) at (210:0.7);
            \coordinate (v225) at (225:0.7);
            \coordinate (v240) at (240:0.7);
            \coordinate (v270) at (270:0.7);
            \coordinate (v300) at (300:0.7);
            \coordinate (v315) at (315:0.7);
            \coordinate (v330) at (330:0.7);
            \draw[tad] (0,0) -- (v90);
            \draw[tad] (0,0) -- (v210);
            \draw[tad] (0,0) -- (v330);
            \fill (v30) circle (1pt);
            \fill (v150) circle (1pt);
            \fill (330:0.35) circle (1pt);
            \fill (0,0) circle (1pt);
            \fill (v90) circle (1pt);
            \fill (v210) circle (1pt);
            \fill (v330) circle (1pt);
          \end{tikzpicture}
        }}
    \right]
    \vcenter{\hbox{
        \begin{tikzpicture}[use Hobby shortcut, scale=0.8]
          \draw (0,0) circle (0.5);
          \draw (135:0.5) .. (90:0.2) .. (45:0.5);
          \draw (180:0.5) -- (180:0.7);
          \draw (0:0.5) -- (0:0.7);
          \fill (180:0.5) circle (1pt);
          \fill (0:0.5) circle (1pt);
          \fill (90:0.5) circle (1pt);
          \fill (45:0.5) circle (1pt);
          \fill (135:0.5) circle (1pt);
        \end{tikzpicture}
      }}
  \right]
  \label{eq:KRg916}\\
  \KRp G_{694} & = \KRp\left[
    \vcenter{\hbox{
        \begin{tikzpicture}[use Hobby shortcut, scale=0.8]
          \draw (0,0) circle (1);
          \coordinate (v0) at (0:1);
          \coordinate (v30) at (30:1);
          \coordinate (v45) at (45:1);
          \coordinate (v60) at (60:1);   
          \coordinate (v90) at (90:1);
          \coordinate (v120) at (120:1);
          \coordinate (v135) at (135:1);
          \coordinate (v150) at (150:1);
          \coordinate (v180) at (180:1);
          \coordinate (v190) at (190:1);
          \coordinate (v210) at (210:1);
          \coordinate (v225) at (225:1);
          \coordinate (v240) at (240:1);
          \coordinate (v270) at (270:1);
          \coordinate (v300) at (300:1);
          \coordinate (v315) at (315:1);
          \coordinate (v330) at (330:1);
          \coordinate (v350) at (350:1);
          \coordinate (u10) at ($(v190)!0.25!(v350)$);
          \coordinate (u20) at ($(v350)!0.25!(v190)$);
          \coordinate (u1) at (u10|-1,-0.5);
          \coordinate (u2) at (u20|-1,-0.5);
          \coordinate (c1) at ($(u2)!0.31!(v150)$);
          \coordinate (c2) at ($(v150)!0.55!(u2)$);
          \draw (v135) .. (90:0.5) .. (v45);
          \draw (v190) .. (u1) .. (u2) .. (v350);
          \draw (u1) -- (v30);
          \draw (u2) -- (c1);
          \draw (v150) -- (c2);
          \draw (v210) -- (210:1.2);
          \draw (v330) -- (330:1.2);
          \draw (v90) -- (90:1.2);
          \fill (u1) circle (1pt);
          \fill (u2) circle (1pt);
          \fill (v90) circle (1pt);
          \fill (v190) circle (1pt);
          \fill (v350) circle (1pt);
          \fill (v45) circle (1pt);
          \fill (v135) circle (1pt);
          \fill (v210) circle (1pt);
          \fill (v330) circle (1pt);
          \fill (v30) circle (1pt);
          \fill (v150) circle (1pt);
        \end{tikzpicture}
      }}
  \right]
  && =
  \mathcal{K}\left[G(1,2+4\ep)
    \vcenter{\hbox{
        \begin{tikzpicture}[use Hobby shortcut, scale=0.8]
          \draw (0,0) circle (0.7);
          \coordinate (v0) at (0:0.7);
          \coordinate (v30) at (30:0.7);
          \coordinate (v45) at (45:0.7);
          \coordinate (v60) at (60:0.7);   
          \coordinate (v90) at (90:0.7);
          \coordinate (v120) at (120:0.7);
          \coordinate (v135) at (135:0.7);
          \coordinate (v150) at (150:0.7);
          \coordinate (v180) at (180:0.7);
          \coordinate (v210) at (210:0.7);
          \coordinate (v225) at (225:0.7);
          \coordinate (v240) at (240:0.7);
          \coordinate (v270) at (270:0.7);
          \coordinate (v300) at (300:0.7);
          \coordinate (v315) at (315:0.7);
          \coordinate (v330) at (330:0.7);
          \coordinate (c1) at ($(v210)!0.43!(v30)$);
          \coordinate (c2) at ($(v30)!0.43!(v210)$);
          \draw (v60) .. (90:0.5) .. (v120);
          \draw (v150) -- (v330);
          \draw (v210) -- (c1);
          \draw (v30) -- (c2);
          \draw (v180) -- (180:1);
          \draw (v0) -- (0:1);
          \fill (v180) circle (1pt);
          \fill (v0) circle (1pt);
          \fill (v60) circle (1pt);
          \fill (v120) circle (1pt);
          \fill (v90) circle (1pt);
          \fill (v30) circle (1pt);
          \fill (v150) circle (1pt);
          \fill (v210) circle (1pt);
          \fill (v330) circle (1pt);
        \end{tikzpicture}
      }}
  \right]
  - \mathcal{K}\left[
    \mathcal{K}\left[
      \vcenter{\hbox{
          \begin{tikzpicture}[use Hobby shortcut, scale=0.8]
            \draw[tad] (0,0) circle (0.5);
            \fill (90:0.5) circle (1pt);
            \fill (210:0.5) circle (1pt);
            \fill (330:0.5) circle (1pt);
          \end{tikzpicture}
        }}
    \right]
    \vcenter{\hbox{
        \begin{tikzpicture}[use Hobby shortcut, scale=0.8]
          \draw (0,0) circle (0.7);
          \coordinate (v0) at (0:0.7);
          \coordinate (v30) at (30:0.7);
          \coordinate (v45) at (45:0.7);
          \coordinate (v60) at (60:0.7);   
          \coordinate (v90) at (90:0.7);
          \coordinate (v120) at (120:0.7);
          \coordinate (v135) at (135:0.7);
          \coordinate (v150) at (150:0.7);
          \coordinate (v180) at (180:0.7);
          \coordinate (v210) at (210:0.7);
          \coordinate (v225) at (225:0.7);
          \coordinate (v240) at (240:0.7);
          \coordinate (v270) at (270:0.7);
          \coordinate (v300) at (300:0.7);
          \coordinate (v315) at (315:0.7);
          \coordinate (v330) at (330:0.7);
          \coordinate (u1) at (-0.3,0);
          \coordinate (u2) at (0.3,0);
          \coordinate (c1) at ($(u1)!0.40!(v60)$);
          \coordinate (c2) at ($(v60)!0.45!(u1)$);
          \draw (160:0.7) .. (u1) .. (u2) .. (20:0.7);
          \draw (u2) -- (v120);
          \draw (u1) -- (c1);
          \draw (v60) -- (c2);
          \draw (v180) -- (180:1);
          \draw (v0) -- (0:1);
          \fill (v180) circle (1pt);
          \fill (v0) circle (1pt);
          \fill (v60) circle (1pt);
          \fill (v120) circle (1pt);
          \fill (v90) circle (1pt);
          \fill (u1) circle (1pt);
          \fill (u2) circle (1pt);
          \fill (160:0.7) circle (1pt);
          \fill (20:0.7) circle (1pt);
        \end{tikzpicture}
      }}
  \right]
  \nonumber
  \\
  &  && -\mathcal{K}\left[\KRp\left[
      \vcenter{\hbox{
          \begin{tikzpicture}[use Hobby shortcut, scale=0.8]
            \draw[tad] (0,0) circle (0.7);
            \coordinate (v0) at (0:0.7);
            \coordinate (v30) at (30:0.7);
            \coordinate (v45) at (45:0.7);
            \coordinate (v60) at (60:0.7);   
            \coordinate (v90) at (90:0.7);
            \coordinate (v120) at (120:0.7);
            \coordinate (v135) at (135:0.7);
            \coordinate (v150) at (150:0.7);
            \coordinate (v180) at (180:0.7);
            \coordinate (v210) at (210:0.7);
            \coordinate (v225) at (225:0.7);
            \coordinate (v240) at (240:0.7);
            \coordinate (v270) at (270:0.7);
            \coordinate (v300) at (300:0.7);
            \coordinate (v315) at (315:0.7);
            \coordinate (v330) at (330:0.7);
            \draw[tad] (0,0) -- (v0);
            \draw[tad] (0,0) -- (v180);
            \draw[tad] (0,0) -- (v270);
            \draw[tad] (v120) .. (0,0.4) .. (v60);
            \fill (0,0) circle (1pt);
            \fill (v90) circle (1pt);
            \fill (v180) circle (1pt);
            \fill (v0) circle (1pt);
            \fill (v270) circle (1pt);
            \fill (v315) circle (1pt);
            \fill (-0.35,0) circle (1pt);
            \fill (v120) circle (1pt);
            \fill (v60) circle (1pt);
          \end{tikzpicture}
        }}
    \right]
    \vcenter{\hbox{
        \begin{tikzpicture}[use Hobby shortcut, scale=0.8]
          \draw (0,0) circle (0.5);
          \draw (180:0.5) -- (180:0.7);
          \draw (0:0.5) -- (0:0.7);
          \fill (180:0.5) circle (1pt);
          \fill (0:0.5) circle (1pt);
          \fill (90:0.5) circle (1pt);
        \end{tikzpicture}
      }}
  \right]                 
  \label{eq:KRg694}\\
  \KRp G_{930} & = \KRp\left[
    \vcenter{\hbox{
        \begin{tikzpicture}[use Hobby shortcut, scale=0.8]
          \draw (0,0) circle (1);
          \coordinate (v0) at (0:1);
          \coordinate (v30) at (30:1);
          \coordinate (v10) at (10:1);
          \coordinate (v45) at (45:1);
          \coordinate (v60) at (60:1);   
          \coordinate (v90) at (90:1);
          \coordinate (v120) at (120:1);
          \coordinate (v135) at (135:1);
          \coordinate (v150) at (150:1);
          \coordinate (v170) at (170:1);
          \coordinate (v180) at (180:1);
          \coordinate (v190) at (190:1);
          \coordinate (v210) at (210:1);
          \coordinate (v225) at (225:1);
          \coordinate (v240) at (240:1);
          \coordinate (v270) at (270:1);
          \coordinate (v300) at (300:1);
          \coordinate (v315) at (315:1);
          \coordinate (v330) at (330:1);
          \coordinate (v350) at (350:1);
          \draw (v120) .. (0,0.7) .. (v60);
          \draw (v150) .. (0,0.3) .. (v30);
          \draw (v170) .. (0,-0.1) .. (v10);
          \draw (v190) .. (0,-0.5) .. (v350);
          \draw (v210) -- (210:1.2);
          \draw (v330) -- (330:1.2);
          \draw (v90) -- (90:1.2);
          \fill (v210) circle (1pt);
          \fill (v330) circle (1pt);
          \fill (v90) circle (1pt);
          \fill (v120) circle (1pt);
          \fill (v60) circle (1pt);
          \fill (v150) circle (1pt);
          \fill (v30) circle (1pt);
          \fill (v170) circle (1pt);
          \fill (v10) circle (1pt);
          \fill (v190) circle (1pt);
          \fill (v350) circle (1pt);
        \end{tikzpicture}
      }}
  \right]
  && =
  \mathcal{K}\left[G(1,2+4\ep)
    \vcenter{\hbox{
        \begin{tikzpicture}[use Hobby shortcut, scale=0.8]
          \draw (0,0) circle (0.7);
          \coordinate (v0) at (0:0.7);
          \coordinate (v30) at (30:0.7);
          \coordinate (v25) at (25:0.7);
          \coordinate (v45) at (45:0.7);
          \coordinate (v60) at (60:0.7);   
          \coordinate (v90) at (90:0.7);
          \coordinate (v120) at (120:0.7);
          \coordinate (v135) at (135:0.7);
          \coordinate (v150) at (150:0.7);
          \coordinate (v155) at (155:0.7);
          \coordinate (v180) at (180:0.7);
          \coordinate (v210) at (210:0.7);
          \coordinate (v225) at (225:0.7);
          \coordinate (v240) at (240:0.7);
          \coordinate (v270) at (270:0.7);
          \coordinate (v300) at (300:0.7);
          \coordinate (v315) at (315:0.7);
          \coordinate (v330) at (330:0.7);
          \draw (v120) .. (90:0.5) .. (v60);
          \draw (v135) .. (90:0.3) .. (v45);
          \draw (v155) .. (90:0.0) .. (v25);
          \draw (v180) -- (180:1);
          \draw (v0) -- (0:1);
          \fill (v180) circle (1pt);
          \fill (v0) circle (1pt);
          \fill (v45) circle (1pt);
          \fill (v135) circle (1pt);
          \fill (v60) circle (1pt);
          \fill (v120) circle (1pt);
          \fill (v90) circle (1pt);
          \fill (v155) circle (1pt);
          \fill (v25) circle (1pt);
        \end{tikzpicture}
      }}
  \right]
  - \mathcal{K}\left[
    \mathcal{K}\left[
      \vcenter{\hbox{
          \begin{tikzpicture}[use Hobby shortcut, scale=0.8]
            \draw[tad] (0,0) circle (0.5);
            \fill (90:0.5) circle (1pt);
            \fill (210:0.5) circle (1pt);
            \fill (330:0.5) circle (1pt);
          \end{tikzpicture}
        }}
    \right]
    \vcenter{\hbox{
        \begin{tikzpicture}[use Hobby shortcut, scale=0.8]
          \draw (0,0) circle (0.7);
          \coordinate (v0) at (0:0.7);
          \coordinate (v30) at (30:0.7);
          \coordinate (v25) at (25:0.7);
          \coordinate (v45) at (45:0.7);
          \coordinate (v60) at (60:0.7);   
          \coordinate (v90) at (90:0.7);
          \coordinate (v120) at (120:0.7);
          \coordinate (v135) at (135:0.7);
          \coordinate (v150) at (150:0.7);
          \coordinate (v155) at (155:0.7);
          \coordinate (v180) at (180:0.7);
          \coordinate (v210) at (210:0.7);
          \coordinate (v225) at (225:0.7);
          \coordinate (v240) at (240:0.7);
          \coordinate (v270) at (270:0.7);
          \coordinate (v300) at (300:0.7);
          \coordinate (v315) at (315:0.7);
          \coordinate (v330) at (330:0.7);
          \draw (v120) .. (90:0.5) .. (v60);
          \draw (v135) .. (90:0.3) .. (v45);
          \draw (v155) .. (90:0.0) .. (v25);
          \draw (v180) -- (180:1);
          \draw (v0) -- (0:1);
          \fill (v180) circle (1pt);
          \fill (v0) circle (1pt);
          \fill (v45) circle (1pt);
          \fill (v135) circle (1pt);
          \fill (v60) circle (1pt);
          \fill (v120) circle (1pt);
          \fill (v90) circle (1pt);
          \fill (v155) circle (1pt);
          \fill (v25) circle (1pt);
        \end{tikzpicture}
      }}
  \right]
  \nonumber
  \\
  &  && - \mathcal{K}\left[
    \KRp\left[
      \vcenter{\hbox{
          \begin{tikzpicture}[use Hobby shortcut, scale=0.8]
            \draw[tad] (0,0) circle (0.5);
            \draw[tad] (180:0.5) -- (0:0.5);
            \fill (90:0.5) circle (1pt);
            \fill (180:0.5) circle (1pt);
            \fill (0:0.5) circle (1pt);
            \fill (210:0.5) circle (1pt);
            \fill (330:0.5) circle (1pt);
          \end{tikzpicture}
        }}
    \right]
    \vcenter{\hbox{
        \begin{tikzpicture}[use Hobby shortcut, scale=0.8]
          \draw (0,0) circle (0.7);
          \coordinate (v0) at (0:0.7);
          \coordinate (v30) at (30:0.7);
          \coordinate (v25) at (25:0.7);
          \coordinate (v45) at (45:0.7);
          \coordinate (v60) at (60:0.7);   
          \coordinate (v90) at (90:0.7);
          \coordinate (v120) at (120:0.7);
          \coordinate (v135) at (135:0.7);
          \coordinate (v150) at (150:0.7);
          \coordinate (v155) at (155:0.7);
          \coordinate (v180) at (180:0.7);
          \coordinate (v210) at (210:0.7);
          \coordinate (v225) at (225:0.7);
          \coordinate (v240) at (240:0.7);
          \coordinate (v270) at (270:0.7);
          \coordinate (v300) at (300:0.7);
          \coordinate (v315) at (315:0.7);
          \coordinate (v330) at (330:0.7);
          \draw (v135) .. (90:0.3) .. (v45);
          \draw (v155) .. (90:0.0) .. (v25);
          \draw (v180) -- (180:1);
          \draw (v0) -- (0:1);
          \fill (v180) circle (1pt);
          \fill (v0) circle (1pt);
          \fill (v45) circle (1pt);
          \fill (v135) circle (1pt);
          \fill (v90) circle (1pt);
          \fill (v155) circle (1pt);
          \fill (v25) circle (1pt);
        \end{tikzpicture}
      }}
  \right]
  - \mathcal{K}\left[\KRp\left[
      \vcenter{\hbox{
          \begin{tikzpicture}[use Hobby shortcut, scale=0.8]
            \draw[tad] (0,0) circle (0.7);
            \coordinate (v0) at (0:0.7);
            \coordinate (v30) at (30:0.7);
            \coordinate (v45) at (45:0.7);
            \coordinate (v60) at (60:0.7);   
            \coordinate (v90) at (90:0.7);
            \coordinate (v120) at (120:0.7);
            \coordinate (v135) at (135:0.7);
            \coordinate (v150) at (150:0.7);
            \coordinate (v180) at (180:0.7);
            \coordinate (v210) at (210:0.7);
            \coordinate (v225) at (225:0.7);
            \coordinate (v240) at (240:0.7);
            \coordinate (v270) at (270:0.7);
            \coordinate (v300) at (300:0.7);
            \coordinate (v315) at (315:0.7);
            \coordinate (v330) at (330:0.7);
            \draw[tad] (v0) -- (v180);
            \draw[tad] (v120) .. (0,0.4) .. (v60);
            \fill (v225) circle (1pt);
            \fill (v315) circle (1pt);
            \fill (v90) circle (1pt);
            \fill (v180) circle (1pt);
            \fill (v60) circle (1pt);
            \fill (v120) circle (1pt);
            \fill (v0) circle (1pt);
          \end{tikzpicture}
        }}
    \right]
    \vcenter{\hbox{
        \begin{tikzpicture}[use Hobby shortcut, scale=0.8]
          \draw (0,0) circle (0.5);
          \draw (135:0.5) .. (90:0.2) .. (45:0.5);
          \draw (180:0.5) -- (180:0.7);
          \draw (0:0.5) -- (0:0.7);
          \fill (180:0.5) circle (1pt);
          \fill (0:0.5) circle (1pt);
          \fill (90:0.5) circle (1pt);
          \fill (45:0.5) circle (1pt);
          \fill (135:0.5) circle (1pt);
        \end{tikzpicture}
      }}
  \right]
  \nonumber
  \\
  &  && - \mathcal{K}\left[\KRp\left[
      \vcenter{\hbox{
          \begin{tikzpicture}[use Hobby shortcut, scale=0.8]
            \draw[tad] (0,0) circle (0.7);
            \coordinate (v0) at (0:0.7);
            \coordinate (v30) at (30:0.7);
            \coordinate (v45) at (45:0.7);
            \coordinate (v60) at (60:0.7);   
            \coordinate (v90) at (90:0.7);
            \coordinate (v120) at (120:0.7);
            \coordinate (v135) at (135:0.7);
            \coordinate (v150) at (150:0.7);
            \coordinate (v180) at (180:0.7);
            \coordinate (v210) at (210:0.7);
            \coordinate (v225) at (225:0.7);
            \coordinate (v240) at (240:0.7);
            \coordinate (v270) at (270:0.7);
            \coordinate (v300) at (300:0.7);
            \coordinate (v315) at (315:0.7);
            \coordinate (v330) at (330:0.7);
            \draw[tad] (v150) .. (0,0.2) .. (v30);
            \draw[tad] (v180) .. (0,-0.1) .. (v0);
            \draw[tad] (v120) .. (0,0.4) .. (v60);
            \fill (v90) circle (1pt);
            \fill (v180) circle (1pt);
            \fill (v0) circle (1pt);
            \fill (v150) circle (1pt);
            \fill (v30) circle (1pt);
            \fill (v120) circle (1pt);
            \fill (v60) circle (1pt);
            \fill (v225) circle (1pt);
            \fill (v315) circle (1pt);
          \end{tikzpicture}
        }}
    \right]
    \vcenter{\hbox{
        \begin{tikzpicture}[use Hobby shortcut, scale=0.8]
          \draw (0,0) circle (0.5);
          \draw (180:0.5) -- (180:0.7);
          \draw (0:0.5) -- (0:0.7);
          \fill (180:0.5) circle (1pt);
          \fill (0:0.5) circle (1pt);
          \fill (90:0.5) circle (1pt);
        \end{tikzpicture}
      }}
  \right]\label{eq:KRg930}
\end{alignat}

Here $G(n_1,n_2)$ means the one-loop two-point function defined as
\begin{equation}
  \label{eq:Gdef}
  G(n_1,n_2) =
  \vcenter{\hbox{
      \begin{tikzpicture}[use Hobby shortcut, scale=0.8]
        \draw (0,0) circle (0.5);
          \draw (180:0.5) -- (180:0.7);
          \draw (0:0.5) -- (0:0.7);
          \fill (180:0.5) circle (1pt);
          \fill (0:0.5) circle (1pt);
          \node[anchor=south] at (90:0.5) {\small $n_1$};
          \node[anchor=north] at (270:0.5) {\small $n_2$};
        \end{tikzpicture}
      }}
    =
    \frac{\Gamma(\frac{d}{2}-n_1)\Gamma(\frac{d}{2}-n_2)\Gamma(n_1+n_2-\frac{d}{2})}{\Gamma(n_1)\Gamma(n_2)\Gamma(d-n_1-n_2)}.
\end{equation}

\bibliography{phi3l5}

\providecommand{\href}[2]{#2}\begingroup\raggedright\begin{thebibliography}{10}

\bibitem{Pelissetto:2000ek}
A.~Pelissetto and E.~Vicari, \emph{{Critical phenomena and renormalization
  group theory}},
  \href{https://doi.org/10.1016/S0370-1573(02)00219-3}{\emph{Phys. Rept.}
  {\bfseries 368} (2002) 549}
  [\href{https://arxiv.org/abs/cond-mat/0012164}{{\ttfamily
  cond-mat/0012164}}].

\bibitem{Vasilev}
A.N.~Vasil'ev, \emph{Quantum field renormalization group in critical behavior
  theory and stochastic dynamics}, Chapman \& Hall/CRC (Apr., 2004).

\bibitem{Kompaniets:2017yct}
M.V.~Kompaniets and E.~Panzer, \emph{{Minimally subtracted six loop
  renormalization of $O(n)$-symmetric $\phi^4$ theory and critical exponents}},
  \href{https://doi.org/10.1103/PhysRevD.96.036016}{\emph{Phys. Rev.}
  {\bfseries D96} (2017) 036016}
  [\href{https://arxiv.org/abs/1705.06483}{{\ttfamily 1705.06483}}].

\bibitem{Fei:2014yja}
L.~Fei, S.~Giombi and I.R.~Klebanov, \emph{{Critical $O(N)$ models in
  $6-\epsilon$ dimensions}},
  \href{https://doi.org/10.1103/PhysRevD.90.025018}{\emph{Phys. Rev.}
  {\bfseries D90} (2014) 025018}
  [\href{https://arxiv.org/abs/1404.1094}{{\ttfamily 1404.1094}}].

\bibitem{Fei:2014xta}
L.~Fei, S.~Giombi, I.R.~Klebanov and G.~Tarnopolsky, \emph{{Three loop analysis
  of the critical $O(N)$ models in $6-\epsilon$ dimensions}},
  \href{https://doi.org/10.1103/PhysRevD.91.045011}{\emph{Phys. Rev.}
  {\bfseries D91} (2015) 045011}
  [\href{https://arxiv.org/abs/1411.1099}{{\ttfamily 1411.1099}}].

\bibitem{Fradkin:1987ks}
E.S.~Fradkin and M.A.~Vasiliev, \emph{{On the Gravitational Interaction of
  Massless Higher Spin Fields}},
  \href{https://doi.org/10.1016/0370-2693(87)91275-5}{\emph{Phys. Lett.}
  {\bfseries B189} (1987) 89}.

\bibitem{Vasiliev:2003ev}
M.A.~Vasiliev, \emph{{Nonlinear equations for symmetric massless higher spin
  fields in (A)dS(d)}},
  \href{https://doi.org/10.1016/S0370-2693(03)00872-4}{\emph{Phys. Lett.}
  {\bfseries B567} (2003) 139}
  [\href{https://arxiv.org/abs/hep-th/0304049}{{\ttfamily hep-th/0304049}}].

\bibitem{Adzhemyan:2019gvv}
L.T.~Adzhemyan, E.V.~Ivanova, M.V.~Kompaniets, A.~Kudlis and A.I.~Sokolov,
  \emph{{Six-loop $\varepsilon$ expansion study of three-dimensional $n$-vector
  model with cubic anisotropy}},
  \href{https://doi.org/10.1016/j.nuclphysb.2019.02.001}{\emph{Nucl. Phys.}
  {\bfseries B940} (2019) 332}
  [\href{https://arxiv.org/abs/1901.02754}{{\ttfamily 1901.02754}}].

\bibitem{Kompaniets:2019xez}
M.V.~Kompaniets, A.~Kudlis and A.I.~Sokolov, \emph{{Six-loop $\epsilon$
  expansion study of three-dimensional $O(n)\times O(m)$ spin models}},
  \href{https://doi.org/10.1016/j.nuclphysb.2019.114874}{\emph{Nucl. Phys.}
  {\bfseries B950} (2020) 114874}
  [\href{https://arxiv.org/abs/1911.01091}{{\ttfamily 1911.01091}}].

\bibitem{Potts:1951rk}
R.B.~Potts, \emph{{Some generalized order - disorder transformations}},
  \href{https://doi.org/10.1017/S0305004100027419}{\emph{Proc. Cambridge Phil.
  Soc.} {\bfseries 48} (1952) 106}.

\bibitem{deAlcantaraBonfim:1980pe}
O.F.~de~Alcantara~Bonfim, J.E.~Kirkham and A.J.~McKane, \emph{{Critical
  Exponents to Order $\epsilon^3$ for $\phi^3$ Models of Critical Phenomena in
  Six $\epsilon$-dimensions}},
  \href{https://doi.org/10.1088/0305-4470/13/7/006}{\emph{J. Phys.} {\bfseries
  A13} (1980) L247}.

\bibitem{deAlcantaraBonfim:1981sy}
O.F.~de~Alcantara~Bonfim, J.E.~Kirkham and A.J.~McKane, \emph{{Critical
  Exponents for the Percolation Problem and the Yang-lee Edge Singularity}},
  \href{https://doi.org/10.1088/0305-4470/14/9/034}{\emph{J. Phys.} {\bfseries
  A14} (1981) 2391}.

\bibitem{Gracey:2015tta}
J.A.~Gracey, \emph{{Four loop renormalization of $\phi^3$ theory in six
  dimensions}}, \href{https://doi.org/10.1103/PhysRevD.92.025012}{\emph{Phys.
  Rev.} {\bfseries D92} (2015) 025012}
  [\href{https://arxiv.org/abs/1506.03357}{{\ttfamily 1506.03357}}].

\bibitem{Batkovich:2016jus}
D.V.~Batkovich, K.G.~Chetyrkin and M.V.~Kompaniets, \emph{{Six loop analytical
  calculation of the field anomalous dimension and the critical exponent $\eta$
  in $O(n)$-symmetric $\varphi^4$ model}},
  \href{https://doi.org/10.1016/j.nuclphysb.2016.03.009}{\emph{Nucl. Phys.}
  {\bfseries B906} (2016) 147}
  [\href{https://arxiv.org/abs/1601.01960}{{\ttfamily 1601.01960}}].

\bibitem{Kompaniets:2016hct}
M.~Kompaniets and E.~Panzer, \emph{{Renormalization group functions of $\phi^4$
  theory in the MS-scheme to six loops}},
  \href{https://doi.org/10.22323/1.260.0038}{\emph{PoS} {\bfseries LL2016}
  (2016) 038} [\href{https://arxiv.org/abs/1606.09210}{{\ttfamily
  1606.09210}}].

\bibitem{Schnetz:2016fhy}
O.~Schnetz, \emph{{Numbers and Functions in Quantum Field Theory}},
  \href{https://doi.org/10.1103/PhysRevD.97.085018}{\emph{Phys. Rev.}
  {\bfseries D97} (2018) 085018}
  [\href{https://arxiv.org/abs/1606.08598}{{\ttfamily 1606.08598}}].

\bibitem{BGS5L}
{Borinsky, M. and Gracey, J. and Kompaniets, M. and Schnetz, O.} to appear.

\bibitem{Baikov:2010hf}
P.A.~Baikov and K.G.~Chetyrkin, \emph{{Four Loop Massless Propagators: An
  Algebraic Evaluation of All Master Integrals}},
  \href{https://doi.org/10.1016/j.nuclphysb.2010.05.004}{\emph{Nucl. Phys.}
  {\bfseries B837} (2010) 186}
  [\href{https://arxiv.org/abs/1004.1153}{{\ttfamily 1004.1153}}].

\bibitem{Lee:2011jt}
R.N.~Lee, A.V.~Smirnov and V.A.~Smirnov, \emph{{Master Integrals for Four-Loop
  Massless Propagators up to Transcendentality Weight Twelve}},
  \href{https://doi.org/10.1016/j.nuclphysb.2011.11.005}{\emph{Nucl. Phys.}
  {\bfseries B856} (2012) 95}
  [\href{https://arxiv.org/abs/1108.0732}{{\ttfamily 1108.0732}}].

\bibitem{Georgoudis:2018olj}
A.~Georgoudis, V.~Goncalves, E.~Panzer and R.~Pereira, \emph{{Five-loop
  massless propagator integrals}},
  \href{https://arxiv.org/abs/1802.00803}{{\ttfamily 1802.00803}}.

\bibitem{ChetyrkinGorishnyLarinTkachov:Analytical5loop}
K.G.~Chetyrkin, S.G.~Gorishny, S.A.~Larin and F.V.~Tkachov, ``{Analiticheskoe
  vychislenie pyatipetlevykh priblizheni\u{i} renormgruppovykh funktsi\u{i}
  modeli $g \varphi^4_{(4)}$ v $\mathrm{MS}$-skheme: podiagrammny\u{i}
  analiz}.'' 1986.

\bibitem{Chetyrkin:1982nn}
K.G.~Chetyrkin and F.V.~Tkachov, \emph{{Infrared {$R$}-operation and
  ultraviolet counterterms in the {MS}-scheme}},
  \href{https://doi.org/10.1016/0370-2693(82)90358-6}{\emph{Phys. Lett.}
  {\bfseries 114B} (1982) 340}.

\bibitem{Chetyrkin:1984xa}
K.G.~Chetyrkin and V.A.~Smirnov, \emph{{R* OPERATION CORRECTED}},
  \href{https://doi.org/10.1016/0370-2693(84)91291-7}{\emph{Phys. Lett.}
  {\bfseries 144B} (1984) 419}.

\bibitem{Chetyrkin:1991mw}
K.G.~Chetyrkin, \emph{{Combinatorics of R, R**(-1), and R* operations and
  asymptotic expansions of Feynman integrals in the limit of large momenta and
  masses}}, .

\bibitem{Chetyrkin:2017ppe}
K.G.~Chetyrkin, \emph{{Combinatorics of $\mathbf{R}$-, $\mathbf{R^{-1}}$-, and
  $\mathbf{R^*}$-operations and asymptotic expansions of feynman integrals in
  the limit of large momenta and masses}},
  \href{https://arxiv.org/abs/1701.08627}{{\ttfamily 1701.08627}}.

\bibitem{Panzer:2013cha}
E.~Panzer, \emph{{On the analytic computation of massless propagators in
  dimensional regularization}},
  \href{https://doi.org/10.1016/j.nuclphysb.2013.05.025}{\emph{Nucl. Phys.}
  {\bfseries B874} (2013) 567}
  [\href{https://arxiv.org/abs/1305.2161}{{\ttfamily 1305.2161}}].

\bibitem{Panzer:2014caa}
E.~Panzer, \emph{{Algorithms for the symbolic integration of hyperlogarithms
  with applications to Feynman integrals}},
  \href{https://doi.org/10.1016/j.cpc.2014.10.019}{\emph{Comput. Phys. Commun.}
  {\bfseries 188} (2015) 148}
  [\href{https://arxiv.org/abs/1403.3385}{{\ttfamily 1403.3385}}].

\bibitem{Baikov:2016tgj}
P.A.~Baikov, K.G.~Chetyrkin and J.H.~Kühn, \emph{{Five-Loop Running of the QCD
  coupling constant}},
  \href{https://doi.org/10.1103/PhysRevLett.118.082002}{\emph{Phys. Rev. Lett.}
  {\bfseries 118} (2017) 082002}
  [\href{https://arxiv.org/abs/1606.08659}{{\ttfamily 1606.08659}}].

\bibitem{Herzog:2017ohr}
F.~Herzog, B.~Ruijl, T.~Ueda, J.A.M.~Vermaseren and A.~Vogt, \emph{{The
  five-loop beta function of Yang-Mills theory with fermions}},
  \href{https://doi.org/10.1007/JHEP02(2017)090}{\emph{JHEP} {\bfseries 02}
  (2017) 090} [\href{https://arxiv.org/abs/1701.01404}{{\ttfamily
  1701.01404}}].

\bibitem{Luthe:2017ttc}
T.~Luthe, A.~Maier, P.~Marquard and Y.~Schroder, \emph{{Complete
  renormalization of QCD at five loops}},
  \href{https://doi.org/10.1007/JHEP03(2017)020}{\emph{JHEP} {\bfseries 03}
  (2017) 020} [\href{https://arxiv.org/abs/1701.07068}{{\ttfamily
  1701.07068}}].

\bibitem{Chetyrkin:2017bjc}
K.G.~Chetyrkin, G.~Falcioni, F.~Herzog and J.A.M.~Vermaseren, \emph{{Five-loop
  renormalisation of QCD in covariant gauges}},
  \href{https://doi.org/10.1007/JHEP12(2017)006, 10.3204/PUBDB-2018-02123,
  10.1007/JHEP10(2017)179}{\emph{JHEP} {\bfseries 10} (2017) 179}
  [\href{https://arxiv.org/abs/1709.08541}{{\ttfamily 1709.08541}}].

\bibitem{Vladimirov:1979zm}
A.A.~Vladimirov, \emph{{Method for Computing Renormalization Group Functions in
  Dimensional Renormalization Scheme}},
  \href{https://doi.org/10.1007/BF01018394}{\emph{Theor. Math. Phys.}
  {\bfseries 43} (1980) 417}.

\bibitem{Tkachov:1981wb}
F.V.~Tkachov, \emph{{A Theorem on Analytical Calculability of Four Loop
  Renormalization Group Functions}},
  \href{https://doi.org/10.1016/0370-2693(81)90288-4}{\emph{Phys. Lett.}
  {\bfseries 100B} (1981) 65}.

\bibitem{Chetyrkin:1981qh}
K.G.~Chetyrkin and F.V.~Tkachov, \emph{{Integration by Parts: The Algorithm to
  Calculate beta Functions in 4 Loops}},
  \href{https://doi.org/10.1016/0550-3213(81)90199-1}{\emph{Nucl. Phys.}
  {\bfseries B192} (1981) 159}.

\bibitem{Chetyrkin:2006dh}
K.G.~Chetyrkin, M.~Faisst, C.~Sturm and M.~Tentyukov, \emph{{epsilon-finite
  basis of master integrals for the integration-by-parts method}},
  \href{https://doi.org/10.1016/j.nuclphysb.2006.02.030}{\emph{Nucl. Phys.}
  {\bfseries B742} (2006) 208}
  [\href{https://arxiv.org/abs/hep-ph/0601165}{{\ttfamily hep-ph/0601165}}].

\bibitem{Tentyukov:1999is}
M.~Tentyukov and J.~Fleischer, \emph{{A Feynman diagram analyzer DIANA}},
  \href{https://doi.org/10.1016/S0010-4655(00)00147-8}{\emph{Comput. Phys.
  Commun.} {\bfseries 132} (2000) 124}
  [\href{https://arxiv.org/abs/hep-ph/9904258}{{\ttfamily hep-ph/9904258}}].

\bibitem{Kuipers:2012rf}
J.~Kuipers, T.~Ueda, J.A.M.~Vermaseren and J.~Vollinga, \emph{{FORM version
  4.0}}, \href{https://doi.org/10.1016/j.cpc.2012.12.028}{\emph{Comput. Phys.
  Commun.} {\bfseries 184} (2013) 1453}
  [\href{https://arxiv.org/abs/1203.6543}{{\ttfamily 1203.6543}}].

\bibitem{Steinhauser:2000ry}
M.~Steinhauser, \emph{{MATAD: A Program package for the computation of MAssive
  TADpoles}},
  \href{https://doi.org/10.1016/S0010-4655(00)00204-6}{\emph{Comput. Phys.
  Commun.} {\bfseries 134} (2001) 335}
  [\href{https://arxiv.org/abs/hep-ph/0009029}{{\ttfamily hep-ph/0009029}}].

\bibitem{Pikelner:2017tgv}
A.~Pikelner, \emph{{FMFT: Fully Massive Four-loop Tadpoles}},
  \href{https://doi.org/10.1016/j.cpc.2017.11.017}{\emph{Comput. Phys. Commun.}
  {\bfseries 224} (2018) 282}
  [\href{https://arxiv.org/abs/1707.01710}{{\ttfamily 1707.01710}}].

\bibitem{Ruijl:2017cxj}
B.~Ruijl, T.~Ueda and J.A.M.~Vermaseren, \emph{{Forcer, a FORM program for the
  parametric reduction of four-loop massless propagator diagrams}},
  \href{https://doi.org/10.1016/j.cpc.2020.107198}{\emph{Comput. Phys. Commun.}
  {\bfseries 253} (2020) 107198}
  [\href{https://arxiv.org/abs/1704.06650}{{\ttfamily 1704.06650}}].

\bibitem{Gorishnii:1989gt}
S.G.~Gorishnii, S.A.~Larin, L.R.~Surguladze and F.V.~Tkachov, \emph{{Mincer:
  Program for Multiloop Calculations in Quantum Field Theory for the
  Schoonschip System}},
  \href{https://doi.org/10.1016/0010-4655(89)90134-3}{\emph{Comput. Phys.
  Commun.} {\bfseries 55} (1989) 381}.

\bibitem{Larin:1991fz}
S.A.~Larin, F.V.~Tkachov and J.A.M.~Vermaseren, \emph{{The FORM version of
  MINCER}}, .

\bibitem{Laporta:2001dd}
S.~Laporta, \emph{{High precision calculation of multiloop Feynman integrals by
  difference equations}}, \href{https://doi.org/10.1016/S0217-751X(00)00215-7,
  10.1142/S0217751X00002157}{\emph{Int. J. Mod. Phys.} {\bfseries A15} (2000)
  5087} [\href{https://arxiv.org/abs/hep-ph/0102033}{{\ttfamily
  hep-ph/0102033}}].

\bibitem{Smirnov:2019qkx}
A.V.~Smirnov and F.S.~Chuharev, \emph{{FIRE6: Feynman Integral REduction with
  Modular Arithmetic}},
  \href{https://doi.org/10.1016/j.cpc.2019.106877}{\emph{Comput. Phys. Commun.}
  {\bfseries 247} (2020) 106877}
  [\href{https://arxiv.org/abs/1901.07808}{{\ttfamily 1901.07808}}].

\bibitem{Lee:2012cn}
R.N.~Lee, \emph{{Presenting LiteRed: a tool for the Loop InTEgrals REDuction}},
   \href{https://arxiv.org/abs/1212.2685}{{\ttfamily 1212.2685}}.

\bibitem{Smirnov:2015mct}
A.V.~Smirnov, \emph{{FIESTA4: Optimized Feynman integral calculations with GPU
  support}}, \href{https://doi.org/10.1016/j.cpc.2016.03.013}{\emph{Comput.
  Phys. Commun.} {\bfseries 204} (2016) 189}
  [\href{https://arxiv.org/abs/1511.03614}{{\ttfamily 1511.03614}}].

\bibitem{2011TMP...169.1450A}
L.T.~{Adzhemyan} and M.V.~{Kompaniets}, \emph{{Renormalization group and the
  {\ensuremath{\epsilon}}-expansion: Representation of the
  {\ensuremath{\beta}}-function and anomalous dimensions by nonsingular
  integrals}},
  \href{https://doi.org/10.1007/s11232-011-0121-z}{\emph{Theoretical and
  Mathematical Physics} {\bfseries 169} (2011) 1450}.

\bibitem{Pismenskii:2015xxg}
A.L.~Pismenskii, \emph{{Calculation of the critical index $\eta$ for the
  $\varphi^3$ theory by the conformal bootstrap method}},
  \href{https://doi.org/10.1007/s11232-015-0360-5}{\emph{Theor. Math. Phys.}
  {\bfseries 185} (2015) 1516}.

\bibitem{Kalagov:2014esa}
G.A.~Kalagov and M.Y.~Nalimov, \emph{{Higher-order asymptotics and critical
  indexes in the $\phi^3$ theory}},
  \href{https://doi.org/10.1016/j.nuclphysb.2014.05.008}{\emph{Nucl. Phys.}
  {\bfseries B884} (2014) 672}.

\bibitem{Vasiliev:1982dc}
A.N.~Vasiliev, Y.M.~Pismak and Y.R.~Khonkonen, \emph{{1/N EXPANSION:
  CALCULATION OF THE EXPONENT ETA IN THE ORDER 1/N**3 BY THE CONFORMAL
  BOOTSTRAP METHOD}}, \href{https://doi.org/10.1007/BF01015292}{\emph{Theor.
  Math. Phys.} {\bfseries 50} (1982) 127}.

\bibitem{Vasiliev:1981dg}
A.N.~Vasiliev, Y.M.~Pismak and Y.R.~Khonkonen, \emph{{1/$N$ Expansion:
  Calculation of the Exponents $\eta$ and Nu in the Order 1/$N^2$ for Arbitrary
  Number of Dimensions}},
  \href{https://doi.org/10.1007/BF01019296}{\emph{Theor. Math. Phys.}
  {\bfseries 47} (1981) 465}.

\bibitem{Broadhurst:1996ur}
D.J.~Broadhurst, J.A.~Gracey and D.~Kreimer, \emph{{Beyond the triangle and
  uniqueness relations: Nonzeta counterterms at large N from positive knots}},
  \href{https://doi.org/10.1007/s002880050500}{\emph{Z. Phys.} {\bfseries C75}
  (1997) 559} [\href{https://arxiv.org/abs/hep-th/9607174}{{\ttfamily
  hep-th/9607174}}].

\bibitem{Gracey:1996ub}
J.A.~Gracey, \emph{{Progress with large N(f) beta functions}},
  \href{https://doi.org/10.1016/S0168-9002(97)00130-7}{\emph{Nucl. Instrum.
  Meth.} {\bfseries A389} (1997) 361}
  [\href{https://arxiv.org/abs/hep-ph/9609409}{{\ttfamily hep-ph/9609409}}].

\bibitem{Eichhorn:2016hdi}
A.~Eichhorn, L.~Janssen and M.M.~Scherer, \emph{{Critical O(N) models above
  four dimensions: Small-N solutions and stability}},
  \href{https://doi.org/10.1103/PhysRevD.93.125021}{\emph{Phys. Rev.}
  {\bfseries D93} (2016) 125021}
  [\href{https://arxiv.org/abs/1604.03561}{{\ttfamily 1604.03561}}].

\end{thebibliography}\endgroup

\end{document}